%% file: main.tex
\documentclass[runningheads, orivec]{llncs}
\usepackage[svgnames,dvipsnames]{xcolor}
\usepackage[T1]{fontenc}
\pdfoutput=1
\usepackage[english]{babel}
\usepackage{csquotes}

\usepackage{hyperref}
\hypersetup{
  colorlinks,
  linkcolor={black},
  citecolor={red!70!black},
  urlcolor={blue!70!black}
}

\usepackage{multirow}
\usepackage{wrapfig}

\usepackage[
sortcites=true,
natbib=true,
]{biblatex}
\usepackage{bbding}

\usepackage[]{graphicx}
\usepackage{array}
\usepackage{tabulary}
\usepackage{multirow}
\usepackage{pifont}
\usepackage{float}
\usepackage{adjustbox}
\usepackage{booktabs}
\usepackage{amsmath}
\usepackage{amssymb}
\usepackage{xstring}
\usepackage{soul}
\usepackage[
  group-minimum-digits = 3,
  group-digits = all,
  group-separator = {\,},
	text-series-to-math = true,
	propagate-math-font = true
]{siunitx}
\usepackage{xspace}
\usepackage{fancyvrb}
\usepackage{balance}

\DeclareSIUnit\wei{Wei}
\DeclareSIUnit\ether{Ether}
\DeclareSIUnit\gas{gas}

\usepackage{tikz}
\usetikzlibrary{shapes,arrows,positioning,shapes.multipart}
\usetikzlibrary{backgrounds}
\usetikzlibrary{fit}
\usetikzlibrary{calc}

\input{incl/circled.tex}
\input{incl/lstconfig.tex}
\input{incl/macros.tex}

\usepackage{cleveref}
\Crefname{lstlisting}{Listing}{Listings}
\newcounter{challenges}
\newenvironment{Challenge}{%
  \refstepcounter{challenges}%
}{}

\crefname{challenge}{challenge}{challenges}

\usepackage{paralist}
\usepackage{silence}
\usepackage[font={small}]{caption}
\usepackage{subfig}

\begin{document}
\date{}

%%%%%%%%%%%%%%%%%%%%%%%%%%%%%%%%%%%%%%%%%%%%%%%%%%%%%%%

\title{HCC: A Language-Independent Hardening Contract Compiler for Smart Contracts} 

%%%%%%%%%%%%%%%%%%%%%%%%%%%%%%%%%%%%%%%%%%%%%%%%%%%%%%%

\author{
    Jens-Rene Giesen\inst{1}\Envelope \and 
    Sebastien Andreina\inst{2} \and
    Michael Rodler\inst{3} \and \\
    Ghassan Karame\inst{4} \and
    Lucas Davi\inst{1}
}
\institute{
    University of Duisburg-Essen, Essen, Germany\\ \email{\{jens-rene.giesen, lucas.davi\}@uni-due.de} \and
    NEC Laboratories Europe\\ \email{sebastien.andreina@neclab.eu} \and
    Amazon Web Services\\ \email{m@mrodler.eu} \and
    Ruhr University Bochum\\ \email{ghassan@karame.org}
}
\authorrunning{Giesen et al.}
\titlerunning{HCC: A Language-Independent Hardening Contract Compiler}

%%%%%%%%%%%%%%%%%%%%%%%%%%%%%%%%%%%%%%%%%%%%%%%%%%%%%%%

\maketitle

%%%%%%%%%%%%%%%%%%%%%%%%%%%%%%%%%%%%%%%%%%%%%%%%%%%%%%%

\begin{abstract}
  \input{abstract}
\end{abstract}

%%%%%%%%%%%%%%%%%%%%%%%%%%%%%%%%%%%%%%%%%%%%%%%%%%%%%%%

\section{Introduction}%
\label{sec:intro}
\input{sections/introduction}

%%%%%%%%%%%%%%%%%%%%%%%%%%%%%%%%%%%%%%%%%%%%%%%%%%%%%%%

\section{Problem Statement \& Related Work}%
\label{sec:background}
\input{sections/problem}

%%%%%%%%%%%%%%%%%%%%%%%%%%%%%%%%%%%%%%%%%%%%%%%%%%%%%%%

\section{HCC: Hardening Contract Compiler}
\label{sec:architecture}
\input{sections/architecture}

%%%%%%%%%%%%%%%%%%%%%%%%%%%%%%%%%%%%%%%%%%%%%%%%%%%%%%%

\section{Application to Ethereum}%
\label{sec:application}
\input{sections/impl-solidity}

%%%%%%%%%%%%%%%%%%%%%%%%%%%%%%%%%%%%%%%%%%%%%%%%%%%%%%%

\section{Evaluation}%
\label{sec:evaluation}
\input{sections/evaluation}

%%%%%%%%%%%%%%%%%%%%%%%%%%%%%%%%%%%%%%%%%%%%%%%%%%%%%%%

\section{Application to Hyperledger Fabric}
\label{sec:fabric}
\input{sections/fabric.tex}

%%%%%%%%%%%%%%%%%%%%%%%%%%%%%%%%%%%%%%%%%%%%%%%%%%%%%%%

\section{Conclusion}%
\label{sec:conclusion}
\input{sections/conclusion}

%%%%%%%%%%%%%%%%%%%%%%%%%%%%%%%%%%%%%%%%%%%%%%%%%%%%%%%

\input{sections/acknowledgements}

%%%%%%%%%%%%%%%%%%%%%%%%%%%%%%%%%%%%%%%%%%%%%%%%%%%%%%%

\printbibliography

%%%%%%%%%%%%%%%%%%%%%%%%%%%%%%%%%%%%%%%%%%%%%%%%%%%%%%%

\end{document}

%% file: incl/circled.tex
\usepackage{tikz}
\usetikzlibrary{shapes,arrows,positioning,shapes.multipart}
\newcommand*\circled[1]{\tikz[baseline=(char.base)]{%
            \node[shape=circle,draw,inner sep=.5pt] (char) {#1};}}

\usepackage{xspace}
\newcommand{\circledd}[1]{\circled{#1}\xspace}

\newcommand{\circleone}{\circledd{1}}
\newcommand{\circletwo}{\circledd{2}}
\newcommand{\circlethree}{\circledd{3}}
\newcommand{\circlefour}{\circledd{4}}
\newcommand{\circlefive}{\circledd{5}}
\newcommand{\circlesix}{\circledd{6}}
\newcommand{\circleseven}{\circledd{7}}
\newcommand{\circleeight}{\circledd{8}}
\newcommand{\circlenine}{\circledd{9}}

%% file: incl/lstconfig.tex
\definecolor{lstgreen}{rgb}{0,0.7,0}
\usepackage{listings}
\lstdefinelanguage{solidity}{%
  morekeywords={new, true, false, function, return,
  switch, var, if, in, while, do, else, case, break,
  contract, mapping, struct, library,
  revert, assert, require, throw, sha3, keccak256,
  sstore, sload, delete},
  morekeywords=[2]{export, public, private, payable, view, returns},
  morekeywords=[3]{bool, uint, int, address, uint256, int256, int8, uint8,
    uint128, int128, bytes, bytes32, bytes4},
  morekeywords=[4]{msg, this, tx},
  sensitive=false,
  comment=[l]{//},
  morecomment=[s]{/*}{*/},
  morestring=[b]',
  morestring=[b]"
}
\lstdefinelanguage{evm}{%
  morekeywords={%
    push1, push2, push3, push32, not, and, or, jump, jumpi, swap1, swap2, push4, jumpdest, add, sub, mul, div, pop, invalid, dup1, dup2, dup3, dup8, stop, lt, iszero, revert, sha3, sload, sstore,
  },
  sensitive=false,
  comment=[l]{//},
  morecomment=[s]{/*}{*/},
}

\definecolor{lstgreen}{rgb}{0,0.6,0}
\lstdefinestyle{plain}{%
  numbers=none,
  frame=none,
  xleftmargin=1pt,
  xrightmargin=1pt,
}

%% file: incl/macros.tex
\newcommand{\toolname}{\textsc{hcc}\xspace}
\newcommand{\sguard}{\textsc{sGuard}\xspace}

\renewcommand{\paragraph}[1]{\noindent\textbf{#1}}

\newcommand{\settikzmark}[1]{%
  \tikz[overlay,remember picture,baseline] \coordinate (#1) at (0,0) {};}

\newcommand{\greenplus}{{\color{green} +}}

\newcommand{\highlight}[2]{%
  \draw[YellowGreen,line width=6pt,opacity=0.2]%
    ([yshift=1.5pt]#1) -- ([yshift=1.5pt]#2);%
}

\newcommand{\tikzlinehighlight}[3]{%
  \draw[#1,line width=6pt,opacity=0.2]%
  ([yshift=1.5pt]#2) -- ([yshift=1.5pt]#3);%
}
\newcommand{\tikzmultilinehighlight}[4]{%
  \node [fill=#1,inner ysep=4pt,fit=(#2) (#3) (#4),opacity=0.2] {};%
}

\newcommand{\diffhighlightadd}[2]{%
  \tikzlinehighlight{YellowGreen}{#1}{#2}
}
\newcommand{\diffhighlightremove}[2]{%
  \tikzlinehighlight{LightCoral}{#1}{#2}
}
\newcommand{\diffhighlightaddmulti}[3]{%
  \tikzmultilinehighlight{YellowGreen}{#1}{#2}{#3}
}

%% file: abstract.tex
Developing secure smart contracts remains a challenging task.
Existing approaches are either impractical or leave the burden to developers for fixing bugs. 
In this paper, we propose the first practical smart contract compiler, called \toolname, which automatically inserts security hardening checks at the source-code level based on a novel and language-independent code property graph (CPG) notation. 
The high expressiveness of our developed CPG allows us to mitigate all of the most common smart contract vulnerabilities, namely reentrancy, integer bugs, suicidal smart contracts, improper use of tx.origin, untrusted delegate-calls, and unchecked low-level call bugs. 
Our large-scale evaluation on 10k real-world contracts and several sets of vulnerable contracts from related work demonstrates that \toolname is highly practical, outperforms state-of-the-art contract hardening techniques, and effectively prevents all verified attack transactions without hampering functional correctness.

%% file: sections/introduction.tex
\noindent Smart contracts are becoming increasingly supported by most modern blockchain platforms.
When used within existing cryptocurrencies such as Ethereum, smart contracts often handle very large amounts of cryptocurrency.
Unfortunately, the supporting software engineering tooling around smart contracts is still in its infancy. Important safety and security features prevalent in modern-day software engineering cycles are either not integrated or are not made sufficiently usable~\cite{sharma_exploring_2022}.

Several high-profile attacks on the Ethereum blockchain, such as the \emph{TheDAO} attack, the \emph{Parity Multisig Wallet} attack, and the more recent \emph{Bacon Protocol} and \emph{Orion Protocol} attacks~\cite{web3-bacon-re,web3-orion-re} have fueled research to strengthen the security of smart contracts.
While the bulk of the research on smart contract security is rooted in formal methods and verification~\cite{Kalra2018zeus,Hildenbrandt2018kevm,Jiao2020soliditysemantics,schneidewind2020ethor}, other proposals approach smart contract security from a testing perspective and utilize symbolic execution~\cite{maian2018,Krupp2018teether} or fuzzing~\cite{wuestholz2020harveyfuzz,efcf2023} to generate inputs and identify bugs and security vulnerabilities.
However, as we will argue in Section~\ref{subsec:codeanalysistooling}, existing solutions are impractical for smart contract developers as they typically are hard to use and suffer from a high false positive and negative rate.
Most importantly, the majority of existing approaches require heavy involvement from software developers for patching vulnerabilities and sorting out false alarms.

On the other hand, existing proposals targeting automated smart contract bytecode rewriting~\cite{smartshield2020,rodler2021} are not able to handle complex attack patterns such as reentrancy. They are also incompatible with the upcoming eWASM bytecode format~\cite{ewasm}.
Further, security checks inserted into the bytecode of a smart contract are specific to one execution environment (e.g., the Ethereum virtual machine), which makes them hardly verifiable by most developers as they do not understand bytecode semantics.
Similarly, contract repair tools operating as source-to-source compilers~\cite{solythesis2020pldis2s,nguyen2021-sguard} either rely on developer-specified invariants, do not support many real-world contracts, or provide limited effectiveness.

\noindent
\paragraph{Contributions.}
In this paper, we propose the first practical smart contract compiler, called \toolname (hardening contract compiler), that automatically inserts security checks at the source-code level.
\toolname ensures full transparency for smart contract developers and is fully compatible with other source code related workflows.
To do so, we develop a novel code-property graph (CPGs)~\cite{yamaguchi2014codeproperty} for smart contracts to detect a wide range of vulnerabilities (\Cref{sec:architecture}).
Our design overcomes a set of unique challenges (\Cref{sub:challenges}), namely:
\begin{inparaenum}[(1)]
    \item it supports inter-procedural and inter-contract analysis,
    \item it accurately determines mitigation points (to avoid false positives),
    \item it efficiently hardens against bugs without breaking legitimate behavior, and
    \item it provides support for a highly dynamic smart contract ecosystem involving both permissioned and permissionless blockchains.
\end{inparaenum}

\noindent
Due to our CPG notation, \toolname can be applied to any smart contract platform and programming language, which is especially important considering the high dynamics in this quite young application domain.
For showcasing our work, our prototype implementation of \toolname targets the Solidity language for Ethereum contracts~(\Cref{sec:application}) which is by far the most popular smart contract language. In addition, we provide a detailed analysis of how \toolname can be applied to Hyperledger Fabric~(\Cref{sec:fabric}), one of the most popular enterprise blockchains. Note that Fabric drastically differs from Ethereum in terms of programming language and blockchain design, allowing us to demonstrate the high expressiveness of our CPG. We are not aware of any approach that is capable of supporting different blockchain platforms.

We demonstrate in detail \toolname's effectiveness on all known patterns of reentrancy and integer bugs, while ensuring that it does not prevent legitimate reentrancy required for flash loans.
These two bug classes are responsible for the most notorious and disastrous attacks observed in existing blockchains.
In addition, \toolname is capable of mitigating suicidal smart contracts, improper use of \texttt{tx.origin}, untrusted \emph{delegatecall} and unchecked low-level call bugs. 

We conducted an extensive evaluation (\Cref{sec:evaluation}) based on a dataset of 10k Solidity smart contracts.
Our dataset also includes vulnerable contracts based on the state-of-the-art datasets~\cite{Ferreira_Torres2018osiris,sereum,yinzhi2020everevolvinggame}.
To comprehensively test for functional correctness, we apply a differential testing method on \num{31629} actual blockchain transactions within these datasets.
Moreover, we experimentally compare \toolname with existing source-level contract repair tools~\cite{nguyen2021-sguard}.
Our experiments show that these do not support many real-world contracts and often fail to protect against real-world exploits (Section~\ref{sec:sguard-comparison}).
Finally, to demonstrate the real-world impact of \toolname, we evaluated it against the 25 top-valued Ethereum contracts with source code available as reported by \href{https://etherscan.io/accounts}{etherscan.io}.
We harden these contracts with \toolname and replay \num{49265} transactions using a differential testing method. None of those transactions failed---demonstrating the high practical value of \toolname. We also evaluate the runtime effects of the mitigations in \Cref{sec:morevulns-eval}.

We also integrated \toolname with Hyperledger Fabric and our results show that \toolname was able to fix all the vulnerabilities reported in the five ERC-20 token contracts from the Osiris~\cite{Ferreira_Torres2018osiris} dataset.
We open source \toolname at \url{https://github.com/uni-due-syssec/hcc}.

%% file: sections/problem.tex
\subsection{Background \& Related Work}
\paragraph{Code Analysis Tooling.}
\label{subsec:codeanalysistooling}
Basic syntax directed analysis usually relies on the abstract syntax tree (AST) and is effective at providing immediate feedback to the developer. 
However, it cannot precisely discover complex security vulnerabilities~\cite{yamaguchi2014codeproperty}.
Various static analysis tools translate smart contract code into a custom Datalog-based intermediate representation (IR)~\cite{Tsankov2018securify, brent2018vandal, grech2019gigahorse, Bose2021sailfish}. 
While these approaches provide good analysis capabilities, they usually do not scale and often do not faithfully model the EVM's semantics~\cite{schneidewind2020soundstaticanalysis}.
Similarly, symbolic execution-based approaches for smart contracts~\cite{Luu2016oyente, mythril, Mossberg2019manticore, Ferreira_Torres2018osiris, Krupp2018teether, ethbmc2020, permenev2020verx, xue-clairevoyance, Torres2021confuzzius} achieve high precision, but suffer from scalability issues in practice.

Further, existing code analysis tools require developers to choose between coverage and precision.
Good coverage often entails many false alarms, bearing the risk that developers simply start to ignore them.
A recent study shows that even experienced developers fail to correctly identify vulnerabilities in smart contracts using such tools~\cite{sharma_exploring_2022}.
On the other hand, tools with a low number of false alarms tend to miss vulnerabilities.
Moreover, the majority of existing code analysis tools focus on reporting potential vulnerabilities, but do not provide any guidance on how to fix and harden the contract.

\paragraph{Contract Hardening.}
Several automated mitigation mechanisms have been proposed~\cite{sereum,Torres2019aegis,chen2020soda,solythesis2020pldis2s,nguyen2021-sguard,Ferreira_Torres2022-mg}.
On-chain solutions~\cite{Torres2019aegis,sereum} either require changes to the blockchain ecosystem or to smart contract deployment.
Bytecode-level analysis~\cite{Luu2016oyente,Mossberg2019manticore,Ferreira_Torres2018osiris,Krupp2018teether,ethbmc2020} and hardening tools~\cite{smartshield2020,rodler2021,Ferreira_Torres2022-mg} suffer from imprecision due to the lack of high-level information (e.g., source types and control flow), which was a major source for false alarms in prior work~\cite{sereum}.
As such, these approaches cannot easily cover bug classes that involve complex control-flows and data-flows such as reentrancy.
Moreover, hardening on the bytecode level~\cite{smartshield2020,rodler2021,Ferreira_Torres2022-mg} does not allow developers to inspect the modified contract.
In contrast, source-to-source compilation approaches~\cite{yu-repair,solythesis2020pldis2s,nguyen2021-sguard} allow developers to inspect the introduced changes.
For instance, Solythesis~\cite{solythesis2020pldis2s} inserts runtime checks into Solidity contracts.
However, this requires manual specification of the invariants. Hence, incorrect or incomplete invariants (i.e., specification bugs) would lead to vulnerable smart contracts.
SCRepair~\cite{yu-repair} and sGuard~\cite{nguyen2021-sguard} both leverage symbolic analysis techniques, thereby suffering from the same scalability issues as code analysis tools discussed above. To tackle this shortcoming, sGuard~\cite{nguyen2021-sguard} develops a custom symbolic analysis.
However, as our analysis in~\Cref{sec:sguard-comparison} shows, sGuard does still not support many complex, real-world contracts and often does not prevent well-known exploits.

\subsection{Challenges}
\label{sub:challenges}
\noindent Contrary to the majority of existing software, it is cumbersome to publish a timely update to a vulnerable smart contract.
In fact, there are a number of challenges that emerge when developing effective and practical hardening compilers for smart contracts, namely:

\vspace{0.5 em}
\begin{Challenge}
\paragraph{Challenge 1: Dynamic Ecosystem.}
\label[challenge]{chal:ecosystems}
Smart contract ecosystems and programming languages are continuously evolving.
There are a number of permissionless and permissioned block\-chain platforms supporting various languages (e.g., Solidity, Go, Java, Rust).
To avoid building a special-purpose compiler for each blockchain platform, our goal is to devise a general-purpose compiler that works independently of the platform specifics and languages.
To date, there does not exist any approach that tackles this challenge.
\end{Challenge}

\vspace{0.5 em}
\begin{Challenge}
\paragraph{Challenge 2: Developer-Friendliness \& False Alarms.}
\label[challenge]{chal:runtime-checks}
Existing approaches require developers to manually investigate each warning to sort out false alarms and fix true vulnerabilities. 
Hence, there is a pressing need for a novel approach to smart contract security that focuses on developer usability and eliminates the necessity for them to acquire domain-specific expertise, particularly in light of the ever-evolving ecosystem (c.f., Challenge~1). 
Moreover, it is imperative that any security solution implemented preserves the integrity of legitimate operations
This calls for the development of security checks that are:
\begin{inparaenum}[(1)]
	\item sufficiently generic to harden against all bug flavors of a specific bug pattern, and
	\item at the same time not too strict to prevent legitimate executions.
\end{inparaenum}
\end{Challenge}

\vspace{0.5 em}
\begin{Challenge}
\paragraph{Challenge 3: Inter-Procedural and Inter-Contract Analysis.}
\label[challenge]{chal:inter-analysis}
Smart contracts suffer from various subtle errors in the program code, e.g., unexpected reentrant calls leading to inconsistent state updates.
Detecting a wide range of complex smart contract bugs necessarily requires inter-procedural and inter-contract analysis.
However, doing so for smart contracts is highly challenging since they can
\begin{inparaenum}[(1)]
  \item create new smart contracts at runtime,
  \item use different types of memory for runtime data and persistent state,
  \item be invoked through different entry functions, and
  \item interact with other contracts, e.g., by invoking public functions of other contracts;
\end{inparaenum}
highlighting the need for a novel graph schema that accurately models all the peculiarities of smart contracts.
\end{Challenge}

\vspace{0.5 em}
\begin{Challenge}
\paragraph{Challenge 4: Determining Mitigation Points.}
\label[challenge]{chal:mitigation-points}
Knowing where to insert security checks is a key requirement for hardening against attacks.
This can be particularly challenging when the bug is, e.g., in a program loop. If the bug is in the loop header for the increment or decrement operation, there is a need to investigate the entire loop to validate whether and how the affected variable is processed inside the loop body in order to avoid missing potential vulnerabilities.
\end{Challenge}

%% file: sections/architecture.tex
\begin{figure}[t]
	\centering
	\includegraphics[width=\linewidth]{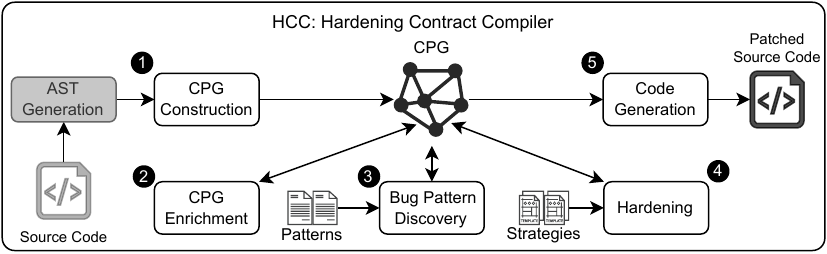}
	\caption{Architecture of \toolname.}%
	\label{fig:architecture}
\vspace{-1.5em}
\end{figure}

To solve the aforementioned challenges, we introduce \toolname, a novel Hardening Contract Compiler that leverages code property graphs (CPGs),
A CPG is a graph model of a program which consists of an AST (abstract syntax tree) enriched with information on control-flows and data-flows. Originally, CPGs have been proposed to detect memory safety bugs in the Linux kernel~\cite{yamaguchi2014codeproperty}.
While memory safety bugs are not relevant for smart contracts---as the latter do not perform low-level memory operations---the superiority  of CPGs in encoding the programs' control-flow and data-flow provides means to protect smart contracts independently of the programming language.

We generate the CPG by instrumenting the compilation process of smart contracts. 
\Cref{fig:architecture} shows the five phases of \toolname's compilation process:
\begin{inparaenum}[(1)]
	\item CPG construction
	\item CPG enrichment,
	\item bug pattern discovery,
	\item hardening, and
	\item code generation.
\end{inparaenum}

First, \toolname transforms the compilers' generated AST into \toolname's novel CPG representation. This step also includes determining the target compiler version to correctly generate matching code (\Cref{chal:ecosystems}).
Second, the \emph{enrichment} phase performs several analysis passes which are especially necessary to determine unresolvable function calls, i.e., calls to other contracts (\Cref{chal:inter-analysis}).
Third, \toolname initiates a vulnerability scan based on CPG graph traversals. To do so, we generated generic CPG patterns to model diverse smart contract bugs (\Cref{chal:inter-analysis,chal:mitigation-points}).
Fourth, we insert hardening patches---again based on generic templates---for insecure code.
Fifth, \toolname emits the hardened AST as source code and continues the ordinary compilation process.

As shown in \Cref{fig:architecture}, \toolname does not remove vulnerabilities but rather adds hardening checks to prevent exploitation. 
The reasoning for this design decision is based on lessons learned from previous code analysis tools that way too often misjudged or missed contract vulnerabilities (see~\Cref{subsec:codeanalysistooling}). 
In doing so, we follow a similar design to the mitigation techniques used in the Desktop world, i.e., control-flow integrity (CFI) assumes that program code will never be free of bugs.  
We note that, in contrast to traditional compiler-based hardening techniques like CFI~\cite{Abadi2005cfi}, we apply our patches to the source code, which allows developers to review the checks.

Note that our modular design enables \toolname to integrate existing detection and patching strategies, while requiring minimal integration efforts.
For yet unknown bug types, additional enrichment steps can be easily included to perform the necessary analysis and CPG extension.
Before we will explain each phase in detail, we introduce our notation.

\input{sections/notation}

\noindent 
\textbf{\ding{182} CPG Construction.} In contrast to an AST, the CPG is a directed graph that consists of diverse types of nodes and directed edges.
Nodes and edges are labeled and can have an arbitrary number of properties.
In \toolname, the AST is the prerequisite for generating the CPG.
Common parser generators such as {ANTLR}~\cite{antlr} support many different programming languages, and enable the creation of compatible ASTs directly from source code for various source languages.
\toolname encodes the sets and elements from \Cref{table:definitions} in the CPG as nodes and edges.
In the CPG, nodes are labeled according to their semantics, e.g., the node representing a function $f\in \mathcal{F}$ is referred to as \texttt{Function} and variables are denoted by the label \texttt{Variable}.
Labeling every piece of data in the CPG effectively creates a schema that facilitates further processing and analysis.
Since \toolname's internal CPG representation preserves the general structure of the original AST, the CPG contains a node for every high-level syntactic element. 
This enables \toolname to insert hardening patches by augmenting the AST.

During CPG construction, \toolname resolves identifiers, e.g., function and variable names, and inserts edges between those identifiers and the variable or function they reference, thereby encoding $a \to b$.
For instance, we insert an edge from a function call expression to the invoked function to encode inter-procedural control-flow information (c.f.~\Cref{chal:inter-analysis}).
This connects function call sites with their underlying implementation to provide inter-procedural control-flow information, which is crucial to tackle \Cref{chal:inter-analysis}.
\toolname encodes the control flow for $\forall s \in {\mathcal{S}}_f$ by creating nodes labeled \emph{Statement}.
Furthermore, \toolname inserts an edge between $f$ and the first statement control-flow-wise in $f$, $s_1 \in {\mathcal{S}}_f$, with the label \emph{body}.
The statements are ordered in ${\mathcal{S}}_f$, so $s_1, s_2 \in {\mathcal{S}}_f$ indicates that $s_1$ precedes $s_2$.
\toolname encodes this control-flow-wise order in the CPG through \emph{leads} edges, i.e., it creates a \emph{leads} edge from $s_1$ to $s_2$.
This facilitates processing and analysis of the control-flow.

\smallskip
\noindent 
\textbf{\ding{183} CPG Enrichment.} Typically, compilers convert the AST into a compiler-specific intermediate representation (IR) which encodes the program's control-flows and data-flows.
While prior work similarly translate the AST into a custom IR~\cite{slither,smartcheck2018,Kalra2018zeus}, \toolname performs control-flow and data-flow analysis directly in the CPG by enriching it with control and data-flow information.
In contrast to IR-based approaches, the CPG allows us to maintain the AST ensuring the possibility for both code analysis and hardening.
\toolname determines unresolvable function calls and built-in API calls to obtain inter-procedural control-flow data (\Cref{chal:inter-analysis}).
It also performs intra-procedural data-flow analysis to detect accesses to variables.
These analysis passes query the CPG and store the results inside the CPG for additional processing.

\smallskip
\noindent 
\textbf{\ding{184} Bug Pattern Discovery.}
The bug pattern discovery phase is a key phase in \toolname. 
Here, we express bug patterns in the form of sub-graphs of the CPG.
The presence of a bug pattern in the CPG indicates that the contract is susceptible to this bug type. 
In practice, \toolname queries the CPG for bug patterns through graph traversals on the CPG.
We refer to distinct matches of a bug pattern in the CPG as a bug instance $v \in {\mathcal{V}}x$.
\toolname adds a node to the CPG for each detected bug.
Depending on the bug type, this node has unique properties and edges to other CPG nodes. The properties include statements affected by the bug and the mitigation point (\Cref{chal:mitigation-points}).
We developed our own static analysis passes that prioritize a low number of missed bugs rather than only patching truly vulnerable code. 
That said, similarly to control-flow integrity approaches~\cite{Abadi2005cfi}, we aim at hardening the contract code at potentially exploitable points rather than relying on a complete removal of vulnerabilities, which is as of today an unreachable goal.

\smallskip
\noindent \textbf{\ding{185} Hardening.} \toolname generates hardening patches for the bug instances discovered in the previous phase.
To do so, it utilizes hardening templates to construct CPG-level patches against vulnerabilities.
The templates are closely related to vulnerability types. 
Hence, we define templates for all individual vulnerability patterns to cover all possible instances $v \in {\mathcal{V}}_x$.
We designed \toolname's automated analysis and hardening such that the potential for accidentally breaking legitimate contract functionality is very low (\Cref{chal:runtime-checks}).
In \Cref{sec:re-hardening,sec:io-hardening,sec:suicidal,sec:txorigin,sec:untrusteddelegate,sec:uncheckedcalls} we provide detailed descriptions of the hardening templates. 
To perform hardening, \toolname queries the CPG for all bug instances and applies the corresponding hardening template.
In doing so, \toolname resolves and inserts missing expressions in the template.
\toolname augments the CPG by inserting nodes to the CPG as required by the template.

\smallskip
\noindent 
\textbf{\ding{186} Code Generation.}
\toolname generates the hardened source code as its last step in the compilation workflow.
\toolname traverses the hardened CPG, i.e., the hardened AST sub-graph of the CPG, to generate the hardened source code, 
taking into account the target compiler version (\Cref{chal:ecosystems}).
The rewritten source code is then ready for further review or verification, compilation, and deployment.

%% file: sections/notation.tex
\paragraph{Notation.}
\label{sub:notation}
\noindent The notation we use throughout this paper are listed in \Cref{table:definitions}.
For a given source code, let $\mathcal{S}$ denote the set of all statements.
A statement $s\in\mathcal{S}$ represents a complete line of code (e.g., until the ``;'' or ``\{'' symbols for languages such as Java and Solidity).
Statements can contain one or multiple expressions $e \in {\mathcal{E}}$ that represent a valid unit of code, $\mathcal{E}$ defined as the set of all expressions.
For example, the following assignment statement $\mathrm{isEven}=(a\%2)==0;$ contains two binary expressions: the equality comparison as the first expression, itself nested with an arithmetic expression performing the modulo operation and four unary expressions: the two constants $0$ and $2$, and the two variables $\mathrm{isEven}$ and $a$.
We use the $\overleftarrow{e}$ (resp. $\overrightarrow{e}$) notation to access the left (resp. right) element of a binary expression $e\in\mathcal{E}$.
By $e^r$, we denote recursive accesses to all sub-expressions of $e$, like $\overleftarrow{e}$ and $\overrightarrow{e}$, until the expression cannot be resolved further.
We denote the set of expressions in a given statement $s$ as ${\mathcal{E}}_s$, and define the top level expression of a statement as $\widehat{e}\in{\mathcal{E}}_s$.
We write ${\mathcal{S}}_{\rightarrow s}$ (resp. ${\mathcal{S}}_{s \rightarrow}$) to designate the former (resp. subsequent) statements of $s\in\mathcal{S}$ in the execution flow.

Since programming languages are based on functions, we define the subset ${\mathcal{S}}_f\subseteq\mathcal{S}$ for statements part of a given function $f$ from the set of all defined functions $\mathcal{F}$.
Functions are further defined through their arguments, $\mathcal{A}_f$, and their local variables, $\mathcal{L}_f$.
We refer to the set of persistent variables (e.g., the database of the smart contract) by $\mathcal{P}$.
Since access and updates of the persistent variables need to be strictly monitored to prevent attacks, we additionally define the set ${\mathcal{U}}=\{s\in {\mathcal{S}}:\exists p\in {\mathcal{P}}~|~s~\mathrm{updates}~p\}$, as the set of all statements $s$ that \emph{update} a persistent variable, and its filtered version ${\mathcal{U}}_p=\{s\in {\mathcal{S}}:s~\mathrm{updates}~p\}$, the set of statements $s\in\mathcal{U}$ that update a specific persistent variable $p\in \mathcal{P}$.
Using the previous notation, we define a smart contract $c$ from the set of all smart contracts $\mathcal{C}$ as the sets of persistent variables $c.\mathcal{P}$, and functions $c.\mathcal{F}$ declared in the contract $c$.
Additionally, we define the helper notations: we write $r\to t,~r\in\mathcal{E}$ to mean $r$ refers to $t$, where $t$ is any variable or function.
We denote the set of binary operations, such as additions, by ${\mathcal{E}}^{BO}$ and the set of binary comparisons, such as equality, by ${\mathcal{E}}^{BC}$.
The set of call expression ${\mathcal{E}}^C$ refers to expressions $e$ referencing a function $f$, i.e., $e\in{\mathcal{E}}_s|e\to f$.
Note that $f$ can be either a defined ($f \in {\mathcal{F}}$) or undefined ($f \notin {\mathcal{F}}$) function.
For $c\in \mathcal{E}^C$, $addr(c)$ refers to the callee contract of $c$.
Finally, we denote by ${\mathcal{V}}_x$  the set of statements vulnerable to vulnerability type $x$.
\begin{wraptable}{O}{0.56\linewidth}
\vspace{-2em}
\centering\footnotesize
\begin{tabular}{{l}m{0.8\linewidth}}
    \toprule
    \textbf{Set} & \textbf{Members} \\
    \midrule
    $\mathcal{C}$                   & All contracts defined in the source code.\\                  
    $\mathcal{P}$                   & Set of all persistent variables. \\  
    $c.\mathcal{P}$                   & Persistent variables declared in contract $c\in \mathcal{C}$. \\           
    $\mathcal{F}$                     & Set of all declared functions.\\                              
    $\mathcal{A}_f$                & Function arguments of $f \in \mathcal{F}$. \\                          
    $\mathcal{L}_f$              & Local variables declared in $f \in \mathcal{F}$.\\                     
    $\mathcal{S}$                     & All statements in the source code.\\                          
    $\mathcal{E}$                     & All expressions in the source code.\\              
    ${\mathcal{V}}_x$                  & All instances of vulnerability $x$.\\                         
    \midrule
    \multicolumn{2}{c}{\emph{Subsets of $\mathcal{S}$}} \\
    \midrule
    ${\mathcal{S}}_f$                   & Statements in function $f \in \mathcal{F}$.\\
    ${\mathcal{S}}_{\rightarrow s}$     & Set of statements former to $s$.\\                            
    ${\mathcal{S}}_{s\rightarrow}$      & Set of statements subsequent to $s$.\\                               
    $\mathcal{U}$                   & Statements $s \in \mathcal{S}$ updating any $p \in \mathcal{P}$.\\           
    ${\mathcal{U}}_p$                   & Statements $s \in \mathcal{S}$ updating $p \in \mathcal{P}$.\\           
    \midrule
    \multicolumn{2}{c}{\emph{Subsets of $\mathcal{E}$}} \\
    \midrule
    ${\mathcal{E}}_s$                   & Expressions in $s \in \mathcal{S}$, ${\mathcal{E}}_s \subseteq \mathcal{E}$. \\                                   
    ${\mathcal{E}}^C$                   & Set of function call expressions.\\                              
    ${\mathcal{E}}^{BO}$                    & Set of binary operation expressions.\\                        
    ${\mathcal{E}}^{BC}$                    & Set of binary comparison expressions.\\                       
    \midrule
    \multicolumn{2}{c}{\emph{Further Notation}} \\
    \midrule
    $a \to b$               & Reference, $a \in \mathcal{E}$ refers to any $b$. \\                   
    $addr(c)$               & The callee contract for any $c \in{\mathcal{E}}^C$.\\                    
    $\widehat{e}$           & $e \in{\mathcal{E}}_s$: the top-level expression of statement $s$.\\  
    $e^r$                   & Recursively resolve all expressions of $e \in\mathcal{E}$. \\      
    \bottomrule
\end{tabular}
\caption{Definition of the notation for \toolname.}%
\label{table:definitions}
\vspace{-2em}
\end{wraptable}

%% file: sections/impl-solidity.tex
\noindent In this section, we describe our implementation of \toolname for Solidity smart contracts in Ethereum.
Ethereum is the most popular smart contract platform, has been intensively investigated in research, and is currently being used by several startups~\cite{polygon,consensys,opensea}.
Reentrancy vulnerabilities are the most exploited bugs in Ethereum and continue to frequently cause losses of millions of USD~\cite{williams_82m_2021,benson_grim_2021,web3-bacon-re,web3-dolomite-re}.
Reentrancy has been extensively investigated in previous research~\cite{sereum,yinzhi2020everevolvinggame,efcf2023,nguyen2021-sguard},
allowing us to use these previous results as ground truth~\cite{rodler2021,Ferreira_Torres2018osiris,sereum,yinzhi2020everevolvinggame} in the evaluation of our prototype.
As reentrancy features a quite complex attack pattern involving interaction between multiple contracts, it is a perfect target to demonstrate the power of our CPG-based approach.

Thanks to its generic CPG model, \toolname is able to mitigate all the major vulnerabilities in Ethereum~\cite{dasp-top10}. 
In \Cref{sec:io-hardening,sec:suicidal,sec:txorigin,sec:untrusteddelegate,sec:uncheckedcalls}, we describe how \toolname covers suicidal smart contracts, improper use of \texttt{tx.origin} for access control, untrusted DelegateCall and unchecked low-level call bugs, as well as integer bugs.

\subsection{Creating CPGs from Solidity ASTs}
\label{sec:solcpg}
\input{sections/cpg-solidity}

\subsection{Reentrancy Hardening}
\label{sec:re-hardening}
\input{sections/re-hardening}

\subsection{Covering additional Vulnerabilities}
\label{sec:other-vulns}
Due to the high expressiveness of our CPG, \toolname also hardens against other vulnerabilities than reentrancy, namely
\begin{inparaenum}[1)]
    \item suicidal smart contracts,
    \item dangerous use of the global variable \texttt{tx.origin} for access control purposes,
    \item performing low-level \emph{delegatecalls} to untrusted addresses, and
    \item unchecked low-level calls.
\end{inparaenum}
\input{sections/vulnerabilities}

\subsection{Integer Bugs}
\label{sec:io-hardening}
\input{sections/io-hardening}

\subsection{Implementation Details}
\label{sec:implementationdetails}
\input{sections/implementation-details}

%% file: sections/cpg-solidity.tex
\begin{wrapfigure}{O}{0.54\linewidth}
\vspace{-2em}
\centering
  \input{figures/contract}
\vspace{-3em}
\end{wrapfigure}

\noindent
\toolname constructs the CPG by calling the Solidity compiler (solc)~\cite{solc}, parsing its JSON-based AST output, and loading the AST into the graph database.
While the AST is loaded into our database, \toolname assigns labels to AST nodes, and adds edges for data-flows and control-flows.
That is, \toolname generates labels and properties to the CPG that represent high-level concepts that are common in smart contracts (e.g., contract, data structure, and function definitions).

\begin{wrapfigure}{O}{0.54\linewidth}
% \vspace{-1em}
	\centering
	\includegraphics[width=\linewidth]{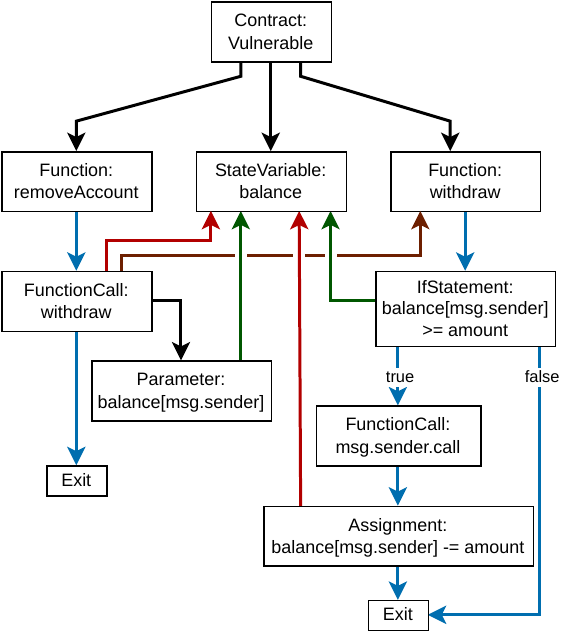}
	\caption{The CPG \toolname constructs for the code from \Cref{lst:solexample}.}%
	\label{fig:basiccpg}
% \vspace{-1em}
\end{wrapfigure}

Consider the smart contract shown in~\Cref{lst:solexample} which serves as an on-chain wallet and as a running example to illustrate every phase of \toolname.
Given sufficient balance, the \textit{withdraw} function allows users to withdraw an amount from the contract.
The \textit{removeAccount} function conveniently allows users to withdraw all their funds.
Without \toolname's hardening (highlighted lines), the \textit{withdraw} function suffers from a reentrancy vulnerability since the balance is updated (Line~15) after an external call (Line~13) which transfers the amount.
\Cref{fig:basiccpg} depicts the CPG that \toolname creates for the original vulnerable contract (prior to hardening).
CPG nodes for \textit{Contract}, \textit{Function} and \textit{StateVariable} are examples of high-level concepts that are common in Solidity smart contracts.
As becomes evident, \toolname utilizes labels to differentiate between the different concepts.

While transforming the AST into the CPG, \toolname already resolves references between identifiers and their entity, i.e., ($a \to b$).
Note that \toolname recognizes different kinds of reference semantics:
\begin{inparaenum}[(1)]
    \item data dependencies (green edges), 
    \item function call references (brown edges), and
    \item \textit{write state} semantics (red edges).
\end{inparaenum}
As shown in \Cref{lst1balcheck}, the condition of the \textit{IfStatement} depends on the \textit{balance} state variable, which is declared in \Cref{lst1defbalance}.
As a result, \toolname inserts the green edge to connect the \textit{IfStatement} and \textit{StateVariable} nodes to represent this data dependency.
Similarly, the function call to the \textit{withdraw} function on \Cref{lst1intcall} refers to the function definition on \Cref{lst1defwithdraw}.
This relationship is represented in \Cref{fig:basiccpg} with the brown edge connecting the \textit{FunctionCall} node from within the \textit{removeAccount} function with the \textit{Function} node.

Lastly, to represent the assignment to the \textit{balance} state variable on \Cref{lst1stateupdate}, \toolname inserts the red edge between the \textit{Assignment} node and the \textit{StateVariable} node to encode the \textit{write state} semantics.
This encodes $b^\prime \to b, b^\prime \in \mathcal{E}, b \in c.\mathcal{P}$, which is subsequently used to construct the set $\mathcal{U}_p$.

However, there are cases where a reference to an entity exists in the AST \textit{before} the entity is declared or defined. 
Whenever \toolname cannot resolve a reference, the reference is stored in a temporary list of unresolved references.
\toolname attempts to resolve the references again as part of its CPG enrichment phase.
Note that during CPG construction, \toolname also inserts control-flow edges (blue edges) between the individual statements of a function to encode $\mathcal{S}_f$.

Control flow constructs such as \textit{IfStatements} or loops are more involved and require special attention to prevent hampering the scalability of \toolname's analyses, e.g., a naive model of control flow for loops would introduce cycles, i.e., ${\mathcal{S}}_{\to s} \cap {\mathcal{S}}_{s \to} \neq \emptyset,~s \in {\mathcal{S}}$.
Hence, for \emph{if} statements, we connect the statement with two different types of edges matching the \texttt{then} branch (\textit{true} label) and \texttt{else} branch (\textit{false} label).
Similarly, we introduce two different kinds of control-flow edges for loops: one that connects the loop header statement with the loop body (true case), and one that connects the last statement of the loop's body with the loop header.

\begin{wrapfigure}{O}{0.5\linewidth}
  \centering
  \includegraphics[width=1\linewidth]{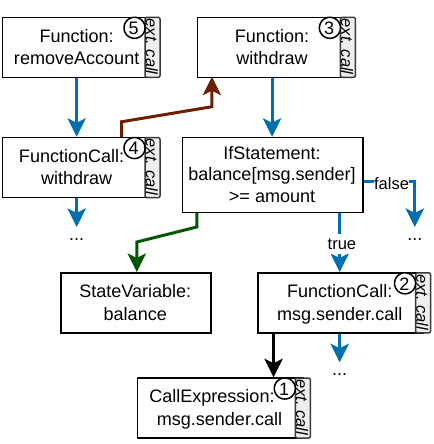}
\caption{Call Analysis during CPG enrichment for \Cref{lst:solexample}.}%
\label{fig:callexample}
 \vspace{-1em}
\end{wrapfigure}

\paragraph{Call Analysis.}
Especially for the case of reentrancy, it is crucial to know the location of external function calls.
Hence, \toolname iterates over all function call expressions $c \in \mathcal{E}^C$ in the CPG and explicitly labels each external call.
External calls are function calls that either target another contract, or statements that utilize low-level functions that are associated with the \emph{address} type of Solidity, i.e., $c \in {\mathcal{E}}^C \land addr(c) \neq {\mathrm{self}} \land c\to f^{\prime} \land f^{\prime} \notin {c.\mathcal{F}}$.
\toolname assumes that control \emph{may} change to any public function of the caller contract.
To detect all cases of external calls (direct or indirect), \toolname applies an analysis consisting of two stages.
Consider again \Cref{lst:solexample} where \Cref{lst1extcall} contains an external call. 
As~\circleone in \Cref{fig:callexample} shows, \toolname adds in the first stage the external call label (\textit{ext. call} label) to the corresponding call expression node in the CPG.
In the second stage, \toolname propagates information about external calls to higher abstraction levels to simplify the bug patterns and hardening strategies, i.e., \toolname labels the statement nodes that contain an external call, as well as the \textit{Function} nodes, whose statements contain external calls.
In our example, \toolname first assigns the \textit{external call} label to the \textit{FunctionCall} node at~\circletwo.
Afterwards, it also labels the \textit{Function} node representing the \textit{withdraw} function (\circlethree).

After this analysis iteration, \toolname can detect external function calls that it could not detect before.
At first sight, the \textit{FunctionCall} at \Cref{lst1intcall} seems like an internal function call, because is calls the \textit{withdraw} function that exists in the same smart contract.
However, after the two stages, \toolname knows that the \textit{withdraw} function performs an external call, so internal function calls to the \textit{withdraw} function perform the external call transitively. 
Fortunately, switching back and forth between the described analysis stages already covers this:
upon executing the first analysis stage again, \toolname now considers the \textit{withdraw} function as \emph{performing an external}.
Hence, it labels the \textit{FunctionCall} node within the \textit{removeAccount} function as an external call at~\circlefour.
During the second analysis stage of the call analysis, \toolname propagates this information further, i.e., to the \textit{Function} node of \textit{removeAccount} (\circlefive).

Similar to external calls, \toolname also identifies \emph{delegate calls} and \emph{constructor calls} to cover the full space of reentrancy attacks~\cite{sereum}.
\toolname identifies the former using the same approach as external calls, but assigns the label \textit{delegate call} in addition to the label \textit{external call}.
For the latter, we analyze Solidity's \emph{new} expressions to check whether the constructor performs an external call.
As the \emph{new} keyword is ambiguous in Solidity (e.g., could be used for new in-memory data types), we need to identify constructor calls that indeed create new contracts and label them as \textit{external calls}.\footnote{Note that creating new contracts by means of the new expression requires the source code of the contract to be created. Hence, \toolname's CPG already contains the contract to be created.}
Prior solutions for reentrancy detection treat constructor calls either as completely trusted or fully untrusted~\cite{sereum}. 
In contrast, HCC can precisely analyze whether a create-based reentrancy is possible since it analyzes the constructor code.

\paragraph{Write State Analysis.}
To find statements that update state, \toolname constructs the set ${\mathcal{U}}$ following a two-step approach similar to the call analysis.
First, \toolname collects all \emph{state updates}--that is, statements that change either state variables or a local variable stored in storage.
Consider \Cref{lst1stateupdate} which updates the \textit{balance} variable.
\toolname labels the \textit{Assignment} node in the CPG as a \textit{state update}, as well as the surrounding \textit{withdraw} function.
Further, \toolname adds a new edge to the CPG (red edge in \Cref{fig:basiccpg}), which directly connects the state update statement with the target state variable, thereby encoding the insertion of $s \in {\mathcal{S}}$ into $\mathcal{U}$.
Moreover, \toolname adds $s$ to the set of statements writing to a state variable, i.e., $s \in {\mathcal{U}}_p, p \in c.{\mathcal{P}}$.
Second, \toolname collects all statements containing internal function calls that update the contract's state and treats these function calls and their surrounding functions just like the original state update statements above.
Further, \toolname includes these function calls in the sets ${\mathcal{U}}$ and ${\mathcal{U}}_p$.
We indicate this in \Cref{fig:basiccpg} with the red edge connecting the \textit{FunctionCall} node of the \textit{removeAccount} function with the \textit{StateVariable} node. 
Similar to the call analysis, \toolname propagates the edges and labels until no new labels were added to the CPG.

%% file: figures/contract.tex
\begin{lstlisting}[language=solidity,escapechar=@,captionpos=b,
label={lst:solexample},
caption={Solidity contract vulnerable to a reentrancy attack. Here, the state update of the \textit{balance} variable happens after the external call of the same function. \toolname inserts the highlighted lines during its hardening phase.}]
contract Vulnerable {
  mapping (address => uint) balance; @\label{lst1defbalance}@
@\settikzmark{hl1s}\greenplus@  mapping(address => bool) balance_lock;@\label{lst1lockdef}\settikzmark{hl1e}@

  @\dots@
  
  function withdraw(uint amount) public { @\label{lst1defwithdraw}@
@\settikzmark{hl3p1}{\color{green}+}\label{lst1lockcheck1}@  require(!balance_lock[msg.sender],
@\settikzmark{hl3p2}@@\greenplus\label{lst1lockcheck2}@  "[HCC] balance_lock[msg.sender] @\dots@");@\settikzmark{hl3p3}@

   if (balance[msg.sender] >= amount) { @\label{lst1balcheck}@
@\settikzmark{hl5s}\greenplus@    balance_lock[msg.sender] = true;@\label{lst1lock}\settikzmark{hl5e}@
     msg.sender.call.value(amount)(""); @\label{lst1extcall}@
@\settikzmark{hl6s}\greenplus@    balance_lock[msg.sender] = false;@\label{lst1unlock}\settikzmark{hl6e}@
     balance[msg.sender] -= amount; @\label{lst1stateupdate}@
    }
  }
  function removeAccount() public { @\label{lst1defremoveaccount}@
   withdraw(balance[msg.sender]); @\label{lst1intcall}@
  }
}
\end{lstlisting}
\begin{tikzpicture}[remember picture, overlay]
\begin{scope}[on background layer]
  \diffhighlightadd{hl1s}{hl1e}
  \diffhighlightaddmulti{hl3p1}{hl3p2}{hl3p3}
  \diffhighlightadd{hl5s}{hl5e}
  \diffhighlightadd{hl6s}{hl6e}
\end{scope}
\end{tikzpicture}

%% file: sections/re-hardening.tex
\noindent 
Reentrancy vulnerabilities enable attackers to reenter a smart contract when the internal state of a smart contract is inconsistent, i.e., when a public function in the victim contract performs an external call to another contract.
Reentrancy bugs are challenging to detect, mostly due to concurrent accesses to state variables in all functions of the smart contract. 
Several classes of reentrancy attacks have been studied in the past~\cite{sereum,efcf2023}:
\begin{inparaenum}[(1)]
	\item same-function reentrancy, in which an attacker maliciously reenters a publicly callable function of the victim contract that operates on an inconsistent state (see \Cref{lst:solexample}),
  \item cross-function reentrancy, where the inconsistent internal state is only exploitable in a sequence of publicly callable functions operating on the same state,
  \item delegate-based reentrancy, which is similar to a same-function reentrancy in which a \emph{delegatecall} replaces either the external call or the state update, and
  \item create-based reentrancy, that happens when the victim contract creates a new contract by calling its constructor before updating the internal state.
\end{inparaenum}
To concisely describe \toolname's reentrancy detection, we define ${\mathcal{D}} = \{c \in {\mathcal{E}}^C: c\to \mathrm{delegatecall}\}$ as the set of low-level delegate calls.

\begin{wrapfigure}{O}{0.5\linewidth}
  \centering
  \includegraphics[width=\linewidth]{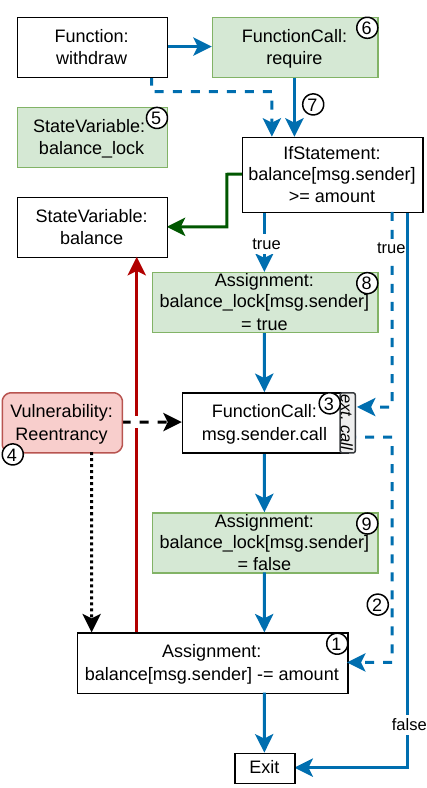}
\caption{Reentrancy Analysis and Hardening for \Cref{lst:solexample}.}%
\label{fig:cpgre}
  \vspace{-3em}
\end{wrapfigure}

\toolname's detection of reentrancy patterns consists of two main steps.
First, \toolname queries the CPG for state updates $u \in {\mathcal{U}}$.
In our example in \Cref{lst:solexample}, a state update happens at \Cref{lst1stateupdate}, which corresponds to the \textit{Assignment} node~\circleone in \Cref{fig:cpgre}.
Second, \toolname queries the CPG for an external call that precedes the state update.
Note that we denote control-flow prior to hardening as dashed blue edges in \Cref{fig:cpgre}.
Starting at the \textit{Assignment} node~\circleone in \Cref{fig:cpgre}, \toolname follows the dashed blue control-flow edge~\circletwo, which in this case directly refers to the external call~\circlethree, i.e.:
\begin{equation*}
  \begin{split}
    {\mathcal{V}}&_{Re} = \{e, u \in {\mathcal{S}}_f:~u\in {\mathcal{U}} \cap {\mathcal{S}}_{e\to}~\land\\
    & \exists~c \in {\mathcal{E}}^C \cap {\mathcal{E}}_s~\land~addr(c) \notin \{\mathrm{self}\}\land\\
    &c\to f^{\prime}~\land~f^{\prime} \notin {\mathcal{F}}\}
  \end{split}
\end{equation*}
${\mathcal{V}}_{Re}$ captures pairs of statements, where the first statement is a function call to an unimplemented function, and the second statement is a state update.
At this point, \toolname already detects the reentrancy vulnerability, as there is a path that leads from an external call to a state update.
Hence, \toolname inserts the red \textit{Vulnerability} node~\circlefour into the CPG for a same-function reentrancy vulnerability.
As \Cref{fig:cpgre} shows, \toolname connects the \textit{Vulnerability} node with the \textit{Assignment} node~\circleone and the \textit{FunctionCall} node~\circlethree.
\toolname uses the \textit{Assignment} during the hardening phase to fully capture all parameters for the mutex locks. 
To detect cross-function reentrancy vulnerabilities $v = (e, u) \in {\mathcal{V}}_{Re}$, \toolname also queries the CPG for \emph{cross-functions} $\mathcal{F}_v^\times$, i.e., functions that operate on the state variable $p \in c.{\mathcal{P}}$:
\begin{equation*}
  \begin{split}
    \mathcal{F}_v^\times = \{ f \in {\mathcal{F}}:~\exists s \in {\mathcal{S}}_f \cap {\mathcal{U}}_p:~\exists p:~u\to p\}
  \end{split}
\end{equation*}
For a given $v \in {\mathcal{V}}_{Re}$, this query returns the set of functions that write to the same state as the vulnerable function containing the reentrancy.
If \emph{cross-functions} exist ($\mathcal{F}_v^\times\neq \varnothing$), \toolname connects the \textit{Vulnerability} node with the corresponding \textit{Function} nodes $f \in \mathcal{F}_v^\times$. 

\noindent
This generic approach covers the same-function and cross-function variants of reentrancy patterns.
Since the call analysis from the CPG enrichment phase also labels \textit{delegate calls} and constructor calls as \textit{external calls}, i.e., ${\mathcal{D}} \subseteq {\mathcal{E}}^C$, this approach also covers delegate-based reentrancy in which a delegate call is performed before a state update.
However, there is a special case ($\mathcal{V}_{Rd}$) in which a \textit{delegate call} is preceded by an \textit{external call}.
In Ethereum, \textit{delegate calls} allow executing the code of the target contract within the calling contracts' context.
As a result, \toolname considers these cases as susceptible to reentrancy:
\begin{equation*}
  \begin{split}
    {\mathcal{V}}_{Rd} = \{&e, d \in {\mathcal{S}}_f:~{\mathcal{D}} \cap {\mathcal{E}}_d \neq \varnothing~\land~d \in {\mathcal{S}}_{e\to}~\land\\
            & \exists c \in {\mathcal{E}}^C \cap {\mathcal{E}}_e~\land~addr(c) \notin \{\mathrm{self}\}~\land\\
            & c\to f^{\prime}~\land~f^{\prime}\notin{\mathcal{F}}\}
  \end{split}
\end{equation*}
Similar to ${\mathcal{V}}_{Re}$, this query returns pairs of statements in order of control-flow with the first one being an external call, but the second statement must contain a delegate call.
However, it is undecidable which state variable--if any--the target contract might write to, so \toolname omits the query for $\mathcal{F}_v^\times$.
\toolname inserts a \textit{Vulnerability} node into the CPG and connects it to the external call inserting a full contract lock for this case.
\Cref{fig:cypherqueryre} in \Cref{sec:cypher} shows how \toolname detects this vulnerability.

In \toolname, the detection of create-based reentrancy (${\mathcal{V}}_{Rn}$) works similar to ${\mathcal{V}}_{Re}$.
However, instead of external calls, \toolname queries for constructor calls which are followed by a state update.
The CPG-based analysis allows \toolname to also analyze the constructor that is being called, even though this constructor is part of another contract, thereby tackling~\Cref{chal:inter-analysis}.
In this way, \toolname only reports a vulnerability if the constructor contains an external call.
For create-based reentrancy, \toolname also conducts the cross-function analysis $\mathcal{F}_v^\times$.
Formally, we express \toolname's detection for create-based reentrancy as follows:
\begin{equation*}
  \begin{split}
    &{\mathcal{V}}_{Rn} = \{n, u \in {\mathcal{S}}_f:~u\in {\mathcal{U}} \cap {\mathcal{S}}_{n\to}~\land~\exists c \in {\mathcal{E}}^C \cap {\mathcal{E}}_n~\land~
            addr(c)~\mathrm{is~undef.}~\land\\
            & c\to f^{\prime}~\land~\exists e \in {\mathcal{S}}_{f^{\prime}}~\land~\exists c^{\prime} \in {\mathcal{E}}^C \cap {\mathcal{E}}_e~\land~
            addr(c^{\prime}) \notin \{\mathrm{self}^{\prime}\}~\land~c^{\prime}\to f^{\prime}~\land~f^{\prime} \notin {\mathcal{F}}\}
  \end{split}
\end{equation*}
While this query is similar to ${\mathcal{V}}_{Re}$, it differs in that it does explicitly look for a function $f \in {\mathcal{F}}$, but with an undefined address.
This is exactly the case for constructors, as there does not exist an address when the constructor is called.
Notably, this pattern only includes sequences where the constructor contains an external call.

\noindent 
For every vulnerability $v \in {\mathcal{V}}_{R} = \{{\mathcal{V}}_{Re} \cup {\mathcal{V}}_{Rd} \cup {\mathcal{V}}_{Rn}\}$, \toolname inserts a vulnerability node into the CPG and labels it according to the vulnerability type, e.g., \emph{create-based reentrancy}.
Finally, \toolname inserts edges between the vulnerability node and the nodes returned from the detection queries.
These edges show which role a node fulfills in the detected bug pattern and are later used by the hardening strategies.

\toolname follows three steps to harden against reentrancy vulnerabilities $(c, u) = v \in \mathcal{V}_R$:
\begin{inparaenum}[(1)]
    \item create a \emph{lock} variable,
    \item insert guard statements that check the locks in the vulnerable function and the \emph{cross-functions}, and
    \item wrap the external call in locking and unlocking statements.
\end{inparaenum}
In our example, \toolname first adds the \textit{StateVariable}~\circlefive node to the CPG in \Cref{fig:cpgre} (\Cref{lst1lockdef} in \Cref{lst:solexample}), i.e., $p^\prime$ for $p: u \to p$.
\toolname considers indices of arrays and mappings, as well as member accesses when constructing \emph{lock} variables.
This avoids false alarms that are present in prior work~\cite{nguyen2021-sguard,sereum}, as those approaches lock the entire contract.

\begin{wrapfigure}{O}{0.7\linewidth}
    \centering
    \input{figures/contract_legitimate_pattern}
\end{wrapfigure}

Consider the contract in~\Cref{lst:legitimate_pattern}.
The \emph{withdrawTo} function contains a reentrancy vulnerability: an attacker contract can set its own address, i.e.,\emph{msg.sender} in this execution context, to the parameter \emph{to} when calling this function.
The attacker contract can then reenter into the \emph{withdrawTo} function and carry out a simple reentrancy attack, leveraging the inconsistent state of the \emph{balance} state variable.
Such an attack is not possible when \emph{msg.sender} and the address \emph{to} are different:
even though a contract deployed at address \emph{to} can reenter this function, the state variable \emph{balance} remains consistent, because the function operates on different indices of the \emph{balance} variable---therefore, this access is legitimate, as it accesses the balances of \emph{different} accounts, which cannot lead to state inconsistency.

Similar to this example, the state variable in our original example from \Cref{lst:solexample} is a \textit{mapping} of \textit{address} to \textit{uint}.
\toolname follows the same access pattern as the affected state variable, which means that the \emph{lock} variable is a \textit{mapping} from \textit{address} to \textit{bool}.
This does not lock the entire contract and allows for cascading withdrawal.

Next, \toolname constructs a guard statement $g$ that checks whether $p^\prime$ is locked, and inserts it at the start of the vulnerable function $f \in \mathcal{F} : u \in \mathcal{S}_f$ so that $s_0 \in S_f = g$.
In \Cref{fig:cpgre}, the green \textit{FunctionCall} node~\circlesix represents $g$.
\toolname also adds a guard statement to the \emph{cross-functions} as they operate on the same state variable, $g = s_0\forall f \in \mathcal{F}_v^\times$.
After node insertion, \toolname adjusts the control-flow of the function to integrate the inserted code.
The \emph{Function} node in \Cref{fig:cpgre} has two outgoing control-flow edges.
The dashed blue edge represents the control-flow before the hardening phase.
After \toolname adds code to the CPG, it redirects the control-flow, so that the solid blue edge and edge~\circleseven represent the control-flow after hardening.

As the last step during hardening, \toolname wraps the \textit{external call} in \emph{lock} ($l$) and \emph{unlock} ($l^\prime$) statements, i.e., \textit{Assignment} statements $l$~\circleeight and $l^\prime$~\circlenine that set the lock to \textit{true} and \textit{false}, respectively.
Similar to the guard statements, \toolname adjusts the control-flow of the function so that the hardening code is executed: $l \in \mathcal{S}_{\to c} \land l^\prime in \mathcal{S}_{c\to}$.

\paragraph{Flash loans.}
Flash loans are a popular mechanism in DeFi, enabling instant loans that do not need any collateral and whose durations span only a single blockchain transaction.
Flash loans rely on legitimate reentrancy, as a borrower must be able to receive and repay the loan within the same transaction.
\Cref{lst:flash} shows a flash loan function:
In \Cref{line:flash_ask_before,line:flash_check_before}, the contract checks that it has enough funds to provide the loan to the borrower.
At \Cref{line:flash_transfer}, the contract transfers the funds to the borrower before calling the borrowers' callback---that is, granting control over the execution to the borrower---in \Cref{line:flash_action}.
Finally, the flash loan contract finishes execution after checking that it has at least the same balance as before the loan plus fees in \Cref{line:flash_ask_after,line:flash_check_after}.

\begin{wrapfigure}{O}{0.7\linewidth}
\input{figures/loan}
\end{wrapfigure}

\noindent
Similar to most other static analysis tools~\cite{Tsankov2018securify, brent2018vandal, grech2019gigahorse, Bose2021sailfish}, \toolname does not prevent this legitimate form of reentrancy.
\toolname detects that locking is not required, as the flash loan functionality does not require the loan provider to update state. 

In case the flash loan contract requires additional DeFi features such as lending and bonding, which should not be executed in the same transaction~\cite{callens2024temporarilyrestrictingsoliditysmart, peapods-audit}, \toolname supports the inclusion of an additional configuration file that specifies sets of mutually exclusive functionalities that users are not allowed to be executed within a single transaction. \toolname then automatically adds a locking mechanism (based on bitmaps) to prevent the unwanted behavior.

%% file: figures/contract_legitimate_pattern.tex
\begin{lstlisting}[language=solidity, escapechar = \%, caption ={Contract vulnerable to a reentrancy attack.}, captionpos = b, label = {lst:legitimate_pattern}]
pragma solidity ^0.4.15;

contract Vulnerable {
  mapping(address => uint256) balance;

  function withdrawTo(uint amount, address to) public {
    if (balance[msg.sender] >= amount) {
      to.call.value(amount)("");
      balance[msg.sender] -= amount;
    }
  }
}
\end{lstlisting}

%% file: figures/loan.tex
\begin{lstlisting}[language=solidity,escapechar=@,captionpos=b, belowskip=-3 em,
label={lst:flash},
caption={A minimal flash loan function.}]
function flashLoan(uint256 amount) external {
  IFlashBorrower r = msg.sender;
  
  @\label{line:flash_ask_before}@uint256 before = t.balanceOf(address(this));
  @\label{line:flash_check_before}@require(before >= amount, "Not enough tokens");
  @\label{line:flash_transfer}@t.transfer(msg.sender, amount);

  @\label{line:flash_action}@require(r.onFlashLoan(@\dots@), "Callback failed");

  @\label{line:flash_ask_after}@uint256 after = t.balanceOf(address(this));
  @\label{line:flash_check_after}@require(after >= before + fee, "Loan not repaid");
}
\end{lstlisting}

%% file: sections/vulnerabilities.tex
\subsubsection{Suicidal Smart Contracts Hardening.}
\label{sec:suicidal}
The global \emph{selfdestruct} function destroys the smart contract. That is, it prevents all further code execution on its address and transfers the balance to an address given as a function parameter.
Suicidal smart contracts suffer from access control bugs that unintentionally allow any entity to call the \emph{selfdestruct} function~\cite{swc-106}.
To prevent exploitation, \toolname captures every call expression to the \emph{selfdestruct} function.
During CPG enrichment, it then queries the CPG for an owner check, i.e., any check comparing addresses to either \emph{msg.sender} or the unsafe \emph{tx.origin}, before the control-flow reaches the call.
To better capture this, we define the set of ownership checks $OC$ as follows:
\begin{equation*}
    \begin{split}
      OC = \{ s& \in {\mathcal{S}}: \exists e \in {\mathcal{E}}_s\cap {\mathcal{E}}^{BC}:~\exists r \in (\overleftarrow{e}~\cup~\overrightarrow{e})^r~\land\\
                  &(r \to \mathrm{tx.origin}~\lor~r \to \mathrm{msg.sender})\}
    \end{split}
\end{equation*}
\begin{figure}[b]
\begin{center}
\begin{minipage}[t]{0.46\linewidth}
  \input{figures/suicidal}
\end{minipage}
\hspace{2ex}
\begin{minipage}[t]{0.5\linewidth}
    \input{figures/txorigin}
\end{minipage}
\end{center}
\end{figure}
$OC$ is the set of all comparison statements where $msg.sender$ or the unsafe $tx.origin$ is used.
Such statements indicate that access control is defined based on the transaction issuer.
\toolname analyzes owner checks during CPG enrichment.
Using this set, we detect the \emph{selfdestruct} vulnerability as follows:
\begin{equation*}
    \begin{split}
      {\mathcal{V}}_{sd} = \{ s &\in {\mathcal{S}}:\exists c \in {\mathcal{E}}^C \cap {\mathcal{E}}_s~\land~c \to \mathrm{selfdestruct}~\land~
            \nexists~o : \{o\in OC~\land~o \in {\mathcal{S}}_{\rightarrow s}\}\}
    \end{split}
\end{equation*}
${\mathcal{V}}_{sd}$ first selects all statements containing a call to \emph{selfdestruct}, and filters out statements preceded by an ownership check.

The contract in \Cref{lst:suicidal} is suicidal because the \emph{destroy} function lacks access control before the call to \emph{selfdestruct} at \Cref{line:selfdestruct}.
To detect this, \toolname collects the statement at \Cref{line:selfdestruct}, and then queries the CPG for a preceding owner check.
As \toolname adds Lines~\ref{line:owner}, \ref{line:ownerset} and \ref{line:oc1} during the hardening phase, \toolname's query for the owner check returns an empty result, i.e., $OC = \varnothing$.
Based on this, \toolname deems the contract as vulnerable at \Cref{line:selfdestruct}.

To mitigate this type of vulnerability, \toolname inserts checks comparing the \emph{owner} of the smart contract with \emph{msg.sender}.
Here, \toolname first queries the CPG for a state variable called \emph{owner} which has an address type.
If this variable does not exist, \toolname will use the variable first declared in the contract with an address type.
However, if none of the state variables have a suitable type (as in our example from \Cref{lst:suicidal}), \toolname will declare an appropriate state variable (\Cref{line:owner}) and set its value to \emph{msg.sender} in the constructor (\Cref{line:ownerset}) of the smart contract.
Once \toolname determines (or creates) the owner variable, \toolname adds an assertion before the call to \emph{selfdestruct} which compares this variable to \emph{msg.sender} (\Cref{line:oc1}).

\subsubsection{Using \texttt{tx.origin} for Access Control.}
\label{sec:txorigin}
Smart contracts can access the account address that initiated the current transaction sequence through the global variable \emph{tx.origin}, which is the bottom-most address in the call stack.
Hence, the \emph{tx.origin} variable should not be used for authorization or access control purposes to avoid impersonation attacks.
\Cref{lst:impersonate} depicts the source code of a smart contract that is susceptible to an impersonation attack (here, the check in \Cref{line:impersonate} relies on \texttt{tx.origin} instead of \texttt{msg.sender}).
A malicious contract, when being called by the owner of this contract, can then execute the \emph{destroy} function granting authorization through \emph{tx.origin}~\cite{swc-115} and cause denial of service.

\toolname detects this vulnerability by checking comparison expressions of addresses against the \emph{tx.origin} global variable.
Namely, \toolname detects this vulnerability as follows:
\begin{equation*}
    \begin{split}
      {\mathcal{V}}&_{to} = \{s \in {\mathcal{S}}:~\exists e \in {\mathcal{E}}_s \cap {\mathcal{E}}^{BC}:~
            \exists r \in (\overleftarrow{e}~\cup~\overrightarrow{e})^r|r \to \mathrm{tx.origin}\}
    \end{split}
\end{equation*}
${\mathcal{V}}_{to}$ matches statements $s$ containing a binary comparison $e \in {\mathcal{E}}^{BC}$, where one of the operands is $tx.origin$.
In our example, this is the assert statement at \Cref{line:impersonate}.
\toolname replaces the variable \emph{tx.origin} in the comparison expression with the variable \emph{msg.sender}, which is always the top-most caller address in the call stack.

\subsubsection{Hardening Untrusted DelegateCalls.}
\label{sec:untrusteddelegate}
The low-level function \emph{delegatecall} allows a smart contract to execute code portions deployed at another address within its own context.
Since the callee code acts on the storage and balance of the caller contract, \emph{delegatecall} into untrusted smart contracts may act maliciously~\cite{swc-112}.
\Cref{lst:untrusted-delegate} shows an example of this vulnerability:
Without the check inserted by \toolname (\Cref{line:delegate-oc}), the DelegateCall in \Cref{line:delegate} executes code from an arbitrary contract.
To detect this vulnerability, \toolname queries the CPG for all \emph{delegatecall} expressions existing in the CPG.
Using the CPG's data-flow information, \toolname queries for the origin of the target contract's address and only trusts addresses from a state variable $p \in c.\mathcal{P}$.
It is impossible to infer the trustworthiness of an unknown call target statically.
Indeed, we argue that this is an access control problem: attackers must be unable to \emph{delegatecall} into malicious contracts, but privileged users are allowed to do so.
For this reason, if the target address originates in a source other than a state variable (e.g., a function parameter or global variable) \toolname also queries the CPG for an owner check in the same way as described for suicidal smart contracts. 
Using the set of delegate calls $\mathcal{D}$ (\Cref{sec:re-hardening}) and the set of ownership checks (\Cref{sec:suicidal}), \toolname detects this vulnerability as follows:
\begin{equation*}
    \begin{split}
      &{\mathcal{V}}_{dc} = \{s \in {\mathcal{S}}_f:~\exists d \in {\mathcal{D}} \cap {\mathcal{E}}_s~|~addr(d) \notin c.{\mathcal{P}}~\land\\
          addr(d) \in \{\mathcal{A}_f&~\cup~\mathcal{L}_f~\cup~\{\mathrm{msg.sender}\} \}~\land~
          \nexists~o : \{o\in OC~\land~o \in {\mathcal{S}}_{s\to}\cup {\mathcal{S}}_{\to s}\}\}
    \end{split}
\end{equation*}
This pattern matches all statements that contain a delegate call and filters out all those statements that are preceded or succeeded by an owner check.
\begin{figure}[b]
\begin{minipage}[t]{0.5\linewidth}
  \input{figures/delegatecall}
\end{minipage}
\hspace{1ex}
\begin{minipage}[t]{0.49\linewidth}
  \input{figures/unchecked}
\end{minipage}
\end{figure}
To mitigate this vulnerability, \toolname adds \Cref{line:delegate-oc} to perform an owner check before the \emph{delegatecall}.

\subsubsection{Unchecked Low-Level Calls.}
\label{sec:uncheckedcalls}
Low-level calls to transfer Ether may result in the execution of another smart contract.
Execution failure of the callee smart contract reverts the Ether transfer and state changes of the callee contract, while the caller resumes execution.
The absence of exception handling can result in locked Ether or denial of service because the caller updates its state despite the transfer being reverted~\cite{swc-113}.
Consider \Cref{lst:unchecked}: the call at \Cref{line:ncall} may fail and \emph{msg.sender} does not receive the funds.
However, without the checks inserted by \toolname (green highlights), the state update at \Cref{line:upd} happens regardless of the calls' success. 
Even worse, this state inconsistency might lead to locked Ethers, i.e., funds that cannot be removed from the smart contract.
\toolname detects this vulnerability by collecting all low-level call expressions from the CPG that are top-level expressions in a statement.
All of them indicate that the call is neither stored in a variable nor checked immediately: 
\begin{equation*}
    \begin{split}
      {\mathcal{V}}&_{ul} = \{ s \in {\mathcal{S}}:~\exists~c \in {\mathcal{E}}^C \cap {\mathcal{E}}_s:~\widehat{c}~\land~addr(c) \notin \{\mathrm{self}\}~\land\\
           &c \to \mathrm{call}~\lor~c \to \mathrm{delegatecall} \}
    \end{split}
\end{equation*}
To mitigate the vulnerability, \toolname wraps the low-level call in a \emph{require} statement, only allowing resumption of the smart contract if the low-level call succeeds (\Cref{line:ncall,line:ncheck}).

%% file: figures/suicidal.tex
\begin{lstlisting}[language=solidity,escapechar=@,captionpos=b, belowskip=-3 em,
label={lst:suicidal},
caption={A suicidal Ethereum smart contract written in Solidity. The highlighted lines show the patches inserted by \toolname.}]
contract Suicidal {
@\label{line:owner}\settikzmark{hl1p1}{\color{OliveGreen}+}@address owner;@\settikzmark{hl1p2}@
@\label{line:ownerset}\settikzmark{hl3p1}{\color{OliveGreen}+}@constructor(){owner=msg.sender;}@\settikzmark{hl3p2}@
 function destroy() public {
@\label{line:oc1}\settikzmark{hl2p1}{\color{OliveGreen}+}@  assert(msg.sender == owner);@\settikzmark{hl2p2}@
@\label{line:selfdestruct}@   selfdestruct(msg.sender);
 }
}
\end{lstlisting}
\begin{tikzpicture}[remember picture, overlay]
  \begin{scope}[on background layer]
  \diffhighlightadd{hl1p1}{hl1p2}
  \diffhighlightadd{hl2p1}{hl2p2}
  \diffhighlightadd{hl3p1}{hl3p2}
\end{scope}
\end{tikzpicture}

%% file: figures/txorigin.tex
\begin{lstlisting}[language=solidity,escapechar=@,captionpos=b, belowskip=-3 em,
label={lst:impersonate},
caption={This function is vulnerable to impersonation attacks. We highlight code inserted (green) and removed (red) by \toolname.}]
function destroy() public {
@\label{line:impersonate}\settikzmark{hl2p1}{\color{OliveGreen} }@ assert(@\settikzmark{hl2p1}{\color{OliveGreen}+}@msg.sender@\settikzmark{hl2p2}@@\settikzmark{hl1p1}{\color{Coral}-}@tx.origin@\settikzmark{hl1p2}@==owner);
@\label{line:privileged}@ selfdestruct(owner);
}
\end{lstlisting}
\begin{tikzpicture}[remember picture, overlay]
  \begin{scope}[on background layer]
  \diffhighlightremove{hl1p1}{hl1p2}
  \diffhighlightadd{hl2p1}{hl2p2}
\end{scope}
\end{tikzpicture}

%% file: figures/delegatecall.tex
\begin{lstlisting}[language=solidity,escapechar=@,captionpos=b, belowskip=-3 em,
label={lst:untrusted-delegate},
caption={This function deliberately executes attacker-controlled code in its own context. We highlight the runtime checks inserted by \toolname in green.}]
function call(address target) public {
@\label{line:delegate-oc}\settikzmark{hl1p1}{\color{OliveGreen}+}@  assert(msg.sender == owner);@\settikzmark{hl1p2}@
@\label{line:delegate}@   target.delegatecall(@\dots@);;
}
\end{lstlisting}
\begin{tikzpicture}[remember picture, overlay]
  \begin{scope}[on background layer]
  \diffhighlightadd{hl1p1}{hl1p2}
\end{scope}
\end{tikzpicture}

%% file: figures/unchecked.tex
\begin{lstlisting}[language=solidity,escapechar=@,captionpos=b, belowskip=-3 em,
label={lst:unchecked},
caption={A function that might lead to state inconsistency. We highlight the runtime checks inserted by \toolname in green.}]
function withdraw(uint val) public {
 require(balance[msg.sender] >= val);
 @\label{line:upd}@balance[msg.sender] -= val;
@\label{line:ncall}@@\settikzmark{hl1p1}{\color{OliveGreen}+}@(bool ok, ) =@\settikzmark{hl1p2}@ msg.sender.call{@\dots@}();
@\label{line:ncheck}\settikzmark{hl2p1}{\color{OliveGreen}+}@require(ok);@\settikzmark{hl2p2}@
}
\end{lstlisting}
\begin{tikzpicture}[remember picture, overlay]
  \begin{scope}[on background layer]
  \diffhighlightadd{hl1p1}{hl1p2}
  \diffhighlightadd{hl2p1}{hl2p2}
\end{scope}
\end{tikzpicture}

%% file: sections/io-hardening.tex
\noindent Integer bugs typically occur when the result of an arithmetic operation grows greater (overflow) or lower (underflow) than the size of the type allows~\cite{swc101}. 
Integer truncation bugs~\cite{cwe-197}, on the other hand, do not result from arithmetic operations but  can occur when an integer value of a certain size is converted into an integer value with a lower size, e.g., when an \emph{uint16} is type-casted to \emph{uint8}.
Our prototype implementation of \toolname hardens against \emph{all} types of integer bugs.

While integer bugs can be prevented using readily available libraries, the handling of those is not straightforward.
For example, the BEC Token contract was actively exploited~\cite{bec-report} due to an integer overflow bug even though the contract already utilized a library for safe integer arithmetic.
We provide a detailed investigation of \toolname's effectiveness against this exploit in \Cref{sub:patch-effectiveness}.

\toolname detects arithmetic bugs by capturing all top-level expressions involving binary arithmetic operations in the CPG:
\begin{equation*}
  {\mathcal{V}}_{ar_1} = \{ s \in {\mathcal{S}} : \exists e \in {\mathcal{E}}^{BO}\}
\end{equation*}
${\mathcal{V}}_{ar_1}$ intuitively matches all statements that contain at least one expression that is a binary operation as potential bugs.
This includes the five standard operations: addition, subtraction, multiplication, exponentiation, and division.
Note that while the division cannot overflow or underflow, its operands could, and hence may require hardening.
To mitigate overflows, \toolname inserts corresponding \emph{assert} statements, which rollback the EVM transaction.
Since arithmetic operations often contain further operations in left and right operands, \toolname first hardens the current expression and then hardens the left and right expression recursively until it reaches a node representing a variable or constant.
As an example, consider the statement $r=\left(a + b\right) * c$.
\toolname will insert a hardening check for all parts of the expression, i.e.,
$assert(a+b \geq b \land \left(a + b\right) * c / \left(a + b\right) = c \lor a+b = 0)$.

\toolname detects truncation bugs by collecting all type conversion expressions from the CPG.
As a helper, we use $\Upsilon$ to denote the set of all type conversion operators.
\begin{equation*}
    \begin{split}
      {\mathcal{V}}_{ar_2} = &\{ s \in {\mathcal{S}} : \exists e \in {\mathcal{E}}_s~\land~e.operator\in \Upsilon\},
    \end{split}
\end{equation*}
To avoid loss of information, \toolname checks whether the value of the cast target stays the same when cast back to its original type.
For example, the type conversion of the variable $uint16~a$ to $uint8(a)$ leads to the following $assert(a==uint16(uint8(a)))$.

%% file: sections/implementation-details.tex
We implement \toolname's CPG on top of the Neo4j~\cite{neo4j} graph database engine, which runs inside the Java Virtual Machine (JVM).
Neo4j provides its own graph query language dubbed Cypher, which we use to implement our enrichment and analysis passes for this prototype. 
Further, as the Neo4j database engine supports in-memory storage only when used in embedded mode, we choose Kotlin~\cite{kotlin} as the language to implement \toolname because of its conciseness.
Our prototype implementation consists of almost \num{4460} lines of Kotlin code for the framework and \num{134} lines of Cypher code.
Note that the most complex query in \toolname, the external call detection during the CPG enrichment phase, consists of 15 lines of Cypher code (see \Cref{fig:cypherquery} below).
On average, queries in \toolname consist of 6--7 lines of code.

\subsubsection{Analysis Queries.}
\label{sec:cypher}
\input{sections/cypher}

%% file: sections/cypher.tex
\noindent
In this section, we describe how two of our queries work.
We point out that this is not the complete set of queries.
Of course, all queries can be found in the source code of \toolname.

\begin{wrapfigure}{L}{0.7\textwidth}
\begin{center}
  \includegraphics[width=0.98\linewidth]{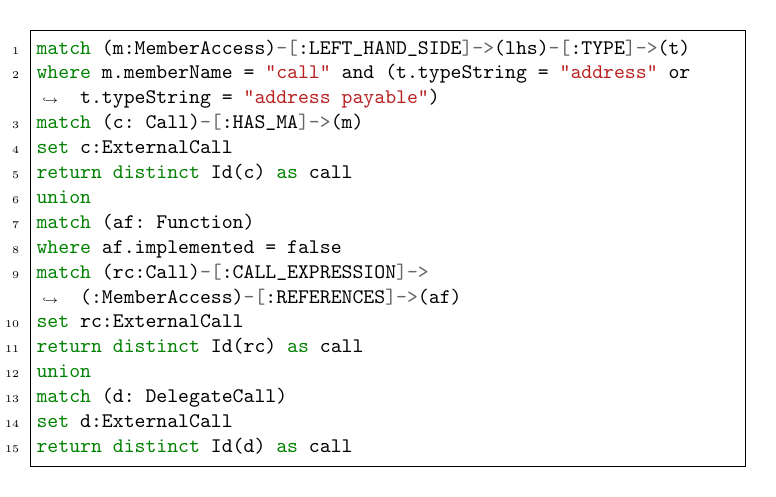}
\end{center}
\caption{Cypher query for the call analysis described in \Cref{sec:solcpg}.}
\label{fig:cypherquery}
\end{wrapfigure}

The query in \Cref{fig:cypherquery} shows the code we implement for the call analysis in \toolname.
From Line~1 to~5, the query searches for the low-level function \textit{call} of the \textit{address} type in Ethereum.
In Line~7 to~11, \toolname collects all function calls in the CPG that have a call target that is \emph{not implemented}, i.e., functions for which only the interface definition is available.
Last, Line~13 and~14 add the \textit{external call} label to all \textit{delegatecalls}.
Due to the \textit{union} directives in Lines~6 and~12 the result set of this query contains all nodes that have received the label so far.
This information is used by \toolname to determine the fix point of this analysis, i.e., \toolname does not proceed the external call analysis once this query yield an empty result set.

\begin{wrapfigure}{O}{0.7\textwidth}
\begin{center}
  \includegraphics[width=0.98\linewidth]{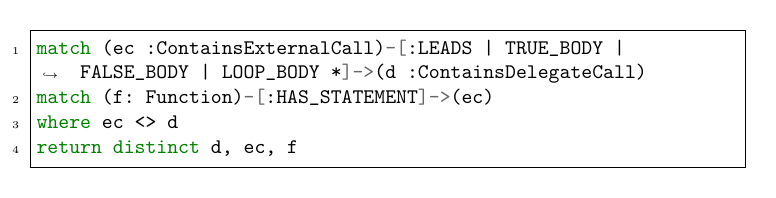}
\end{center}
\caption{Cypher query for the detection of delegate-based reentrancy, as briefly described in \Cref{sec:re-hardening}.}
\label{fig:cypherqueryre}
\end{wrapfigure}

The query in \Cref{fig:cypherqueryre} shows the code we implement for the detection of the special case for delegate-based reentrancy in \toolname.
Remember that in the special case of delegate-based reentrancy, the \textit{delegatecall} takes the role of the state update.
Hence, Line~1 queries for a control-flow between an external call a \textit{delegatecall}.
However, as \toolname also labels \textit{delegatecalls} as external call, Line~3 is required to ensure that \toolname does not assume the exact same call to be both, the external call and the state update at the same time.
Line~3 queries the CPG for the vulnerable function, which \toolname uses when inserting the vulnerability node, and ultimately, during hardening.

%% file: sections/evaluation.tex
\noindent We evaluate the application of \toolname to Solidity smart contracts based on functional correctness and effectiveness.
The former is vital as to ensure that contracts still perform as expected. 
For the latter, we focus on reentrancy and integer bugs to compare \toolname's effectiveness to previous work~\cite{rodler2021,Ferreira_Torres2018osiris,sereum,yinzhi2020everevolvinggame}.
Lastly, we evaluate the runtime effects of \toolname's hardenings for vulnerabilities described in \Cref{sec:suicidal} to~\ref{sec:uncheckedcalls}.

\paragraph{Methodology.}
We utilize the same differential testing approach as in~\cite{rodler2021} based on a large set of important real-world contracts.
To do so, we run a modified Ethereum node based on \emph{geth} to compare execution traces of transactions using both the original and the hardened bytecode.
To measure functional correctness, we ensure that both transaction traces contain the same state updates (i.e., storage writes), ether transfers, and external calls (i.e., any interaction with the blockchain state) in the same order.
We re-execute transactions targeting the smart contract which were successful on the original unhardened contract code. 
This is due to the fact that
\begin{inparaenum}[(1)]
    \item we cannot determine functional correctness of failing transactions, and
    \item \toolname aims at not changing the functional behavior, hence it does not affect the conditions that lead to the failure of transactions in the original contract.
\end{inparaenum}
We evaluate \toolname on contracts that fulfill the following criteria:
\begin{inparaenum}[(1)]
    \item the source code of the contract is available
    \item the contract is programmed in Solidity,
    \item the source code is compatible to a \emph{solc} version of 0.4.12 or above and
    \item \emph{solc} compiles the original source code without errors (and also without the \emph{stack too deep} issue in the Solidity compiler when adding new local variables~\cite{stacktoodeep}).
\end{inparaenum}

\subsection{Functional Correctness}%
\label{sub:functional-correctness}

\noindent
We collected the top 25 high-value accounts with balances ranging from \num{20000} to 7.2~million Ether in August 2022 that fulfill the above criteria. 
We first harden the contracts against exploits that target reentrancy and integer bugs. 
Next, we apply differential testing to determine whether all tested transactions work as intended. We excluded deployment transactions as they cannot be replayed since contracts have already been deployed.
Further, we measure the gas overhead induced by the hardening checks.

\noindent
\Cref{tab:top-results} shows the results of our functional evaluation.
Replaying \num{49265} transactions required one week execution and none of the transactions failed, i.e., \toolname preserves functional correctness for all the top 25 Ethereum contracts.
Notice the varying gas overhead across different contracts. 
This is because
\begin{inparaenum}[(1)]
  \item even slight differences in compiler versions and parameters between the original smart contract and the hardened smart contract can lead to lower gas cost through compiler optimization,
  \item the number of inserted runtime checks varies across contracts,
  \item not all transactions trigger a runtime check.
\end{inparaenum}
As some contracts in Table~\ref{tab:top-results} have a negative value in the minimum gas overhead, we measured the effect of using different compiler versions for patched and unpatched contracts in a separate experiment. 
Our results show that the gas overhead deviates only minimally (between \SI{-0.24}{\percent} and \SI{+0.52}{\percent}) when measurements are done across different compiler versions---which also explains the negative values in~\Cref{tab:top-results}. 
Indeed, using different versions of the compiler before and after hardening can result in lower gas costs.

\begin{figure}[tbp]
	\begin{lstlisting}[language=solidity, escapechar = !,belowskip=-3 em, caption ={The vulnerable function of the exploited BEC Token contract. Highlighted lines mark the runtime checks added by \toolname.}, captionpos = b, label = {listing:bec-hard}]
function batchTransfer(address[]  receivers,
                                        uint  value)
                 public whenNotPaused returns (bool) {
  uint256 cnt = receivers.length;
!\settikzmark{patch1s}\color{OliveGreen}{+}! assert((uint256(cnt) == 0 ) ||!\settikzmark{patch1e}!
!\settikzmark{patch1s2}\color{OliveGreen}{+}!   (uint256(cnt) * value / uint256(cnt) == value));!\settikzmark{patch1e2}!
  uint256 amount = uint256(cnt) * value;
  require(cnt > 0  && cnt <= 20 );
  require(value > 0  && balance[msg.sender] >= amount);
  !\dots
}
\end{lstlisting}
  \begin{tikzpicture}[remember picture, overlay]
    \highlight{patch1s}{patch1e}
    \highlight{patch1s2}{patch1e2}
  \end{tikzpicture}
\end{figure}
We note that all contracts where \toolname inserts reentrancy hardening, i.e., where the RE checks column is not zero in \Cref{tab:top-results}, are variants and extensions of multi-signature wallet contracts~\cite{gnosis}. 
As such, they share large parts of their code, so that \toolname inserts a similar number of reentrancy and integer bug runtime checks.
We find the gas overhead for these contracts largely tolerable. 
For instance, contract 0x2265\ldots1e41, which exhibits the highest overhead, has a mean and maximum gas increase of \num{35149} and \num{99916}. 
This is equivalent to a mean increase of 0.0012 Ether and a maximum increase of 0.0034 Ether (avg. of \SI{34}{\giga\wei} per gas). 
We point out that, in a non-reentrant transaction, it costs \num{10000} gas to store/remove each lock flag (storage writes). 
With the introduction of transient storage operations (TSOs)~\cite{eip-1153}, relying on transient storage could reduce the overhead of locking by 99\% as it would require only \num{100} gas to store/remove the lock.
However, since the majority of the contracts in our dataset predate TSO, we opted not to rely on TSOs. 

Further, our results show that integer bug mitigation results in a very low overhead ranging from \SI{0.14}{\percent} to \SI{10.8}{\percent} with an average of \SI{2.82}{\percent}.
In comparison, the gas overhead of reentrancy mitigation is typically higher due to the storage writes. 
Nonetheless, this overhead is largely tolerable.
\begin{table*}[tbp]
    \centering
    {
      \begin{adjustbox}{max width = \linewidth}
                {\begin{tabular}{@{}l|rr|rrrrrr|rr|r@{}}
            \toprule
            \multirow{2}{*}{Contract}  & \multicolumn{2}{c|}{\# Transactions}          & \multicolumn{6}{c|}{Gas Overhead (gas)}                           & \multicolumn{2}{c|}{Checks}  & \multicolumn{1}{c}{Compile}\\\cline{2-12}
                                       & Total       & Success           & mean        & SD             & mean (\%)              & min                  & max                & median            & RE & IO & \multicolumn{1}{c}{Time (s)}  \\
            \midrule
            \texttt{0x6ba3\ldots{c351}} & \num{4789}    & \num{4789}     & \num{17503} & \num{20978}    & \SI{15.88}\%         & \num{3}              & \num{43777}        & \num{3}           & 3  & 14  & \num{7.97} \\
            \texttt{0x2265\ldots{1e41}} & \num{60}      & \num{60}       & \num{35149} & \num{19144}    & \SI{32.75}\%         & \num{3}              & \num{99916}        & \num{41945}       & 2  & 14  & \num{7.79}\\
            \texttt{0x3d92\ldots{7c18}} & \num{1783}    & \num{1783}     & \num{29979} & \num{19920}    & \SI{26.61}\%         & \num{3}              & \num{43937}        & \num{43103}       & 2  & 14  & \num{7.50}\\
            \texttt{0x51fd\ldots{7ee2}} & \num{630}     & \num{630}      & \num{12111} & \num{10833}    & \SI{9.85}\%          & \num{-128}           & \num{23150}        & \num{21119}       & 1  & 14  & \num{8.04}\\
            \texttt{0x74fe\ldots{4af7}} & \num{159}     & \num{159}      & \num{17806} & \num{8127}     & \SI{20.45}\%         & \num{-56}            & \num{22708}        & \num{21417}       & 1  & 14  & \num{7.15}\\
            \texttt{0x3262\ldots{658a}} & \num{9}      & \num{9}        & \num{24027} & \num{21490}    & \SI{21.92}\%         & \num{3}              & \num{43937}        & \num{42975}       & 2  & 14  & \num{7.55}\\
            \texttt{0xb8d2\ldots{84d0}} & \num{6055}    & \num{6055}     & \num{6166}  & \num{14874}    & \SI{5.41}\%          & \num{0}              & \num{94698}        & \num{3}           & 2  & 14  & \num{7.82}\\
            \texttt{0x7da8\ldots{6cf9}} & \num{384}     & \num{384}      & \num{106}   & \num{407}      & \SI{0.14}\%          & \num{-493}           & \num{2603}         & \num{-63}         & 0  & 10  & \num{6.79}\\
            \texttt{0xc61b\ldots{193c}} & \num{22}      & \num{22}       & \num{0}     & \num{0}        & \SI{0.0}\%           & \num{0}              & \num{0}            & \num{0}           & 0  & 0   & \num{3.93}\\
            \texttt{0x67b6\ldots{9c7a}} & \num{11888}   & \num{11888}    & \num{4824}  & \num{1436}     & \SI{2.62}\%          & \num{0}              & \num{9537}         & \num{4494}        & 0  & 4   & \num{12.53}\\
            \texttt{0xa646\ldots{0a2b}} & \num{160}     & \num{160}      & \num{7183}  & \num{24829}    & \SI{10.8}\%          & \num{-488}           & \num{114680}       & \num{276}         & 0  & 14  & \num{6.89}\\
            \texttt{0x3d45\ldots{1cd1}} & \num{44}      & \num{44}       & \num{327}   & \num{311}      & \SI{0.29}\%          & \num{0}              & \num{994}          & \num{175}         & 0  & 14  & \num{6.87}\\
            \texttt{0x8dc2\ldots{9a20}} & \num{68}      & \num{68}       & \num{0}     & \num{0}        & \SI{0.0}\%           & \num{0}              & \num{0}            & \num{0}           & 0  & 0   & \num{3.92}\\
            \texttt{0x0000\ldots{05fa}} & \num{11016}   & \num{11016}    & \num{6804}  & \num{18481}    & \SI{7.91}\%          & \num{0}              & \num{475580}       & \num{4751}        & 0  & 8   & \num{5.66}\\
            \texttt{0x674b\ldots{bd3f}} & \num{39}      & \num{39}       & \num{174}   & \num{22}       & \SI{0.18}\%          & \num{170}            & \num{314}          & \num{170}         & 0  & 16  & \num{10.27}\\
            \texttt{0x2134\ldots{d469}} & \num{98}      & \num{98}       & \num{716}   & \num{993}      & \SI{1.02}\%          & \num{-65}            & \num{3426}         & \num{358}         & 0  & 9   & \num{7.68}\\
            \texttt{0xc02a\ldots{6cc2}} & \num{9861}    & \num{9861}     & \num{285}   & \num{205}      & \SI{0.88}\%          & \num{33}             & \num{834}          & \num{399}         & 0  & 5   & \num{5.28}\\
            \texttt{0x23ea\ldots{062f}} & \num{1448}    & \num{1448}     & \num{699}   & \num{1518}     & \SI{1.6}\%           & \num{-27745}         & \num{788}          & \num{788}         & 0  & 1   & \num{5.17}\\
            \texttt{0xa646\ldots{0a2b}} & \num{160}     & \num{160}      & \num{7183}  & \num{24829}    & \SI{10.80}\%         & \num{-488}           & \num{114680}       & \num{276}         & 0  & 14  & \num{6.89}\\
            \texttt{0xcafe\ldots{5f2c}} & \num{752}     & \num{752}      & \num{1363}  & \num{9348}     & \SI{1.124}\%         & \num{-52937}         & \num{150697}       & \num{87}          & 0  & 14  & \num{6.94}\\
            \midrule
          Total/\textbf{Average}        & \num{49265}   & \num{49265}    & \textbf{\num{9718}} &        & \textbf{\SI{9.90}\%} & \textbf{\num{-4737}} & \textbf{\num{62312}}  & \textbf{\num{9113}}& 13 & 207 & \textbf{\num{7.13}} \\
            \bottomrule
        \end{tabular}}
      \end{adjustbox}
    }
    \vspace{1ex}
    \caption{Top-25 evaluated Ethereum contracts. RE/IO stand for inserted runtime checks for reentrancy hardening and integer overflow patching.}
    \label{tab:top-results}
\end{table*}

\subsection{Hardening Effectiveness}%
\label{sub:patch-effectiveness}

\noindent To evaluate the hardening effectiveness of \toolname, we leverage existing datasets that contain smart contracts with known vulnerabilities~\cite{Ferreira_Torres2018osiris, sereum, yinzhi2020everevolvinggame} and harden the respective contracts with \toolname.
We evaluate the effectiveness of the two hardening mechanisms separately.
Hence, we construct two distinct datasets, one for each vulnerability class, and apply the differential testing approach described above.

\paragraph{Integer Bug Hardening.}
We evaluate the effectiveness of integer bug hardening on five ERC-20 token contracts from the Osiris~\cite{Ferreira_Torres2018osiris} dataset, which all have confirmed vulnerabilities that have been actively exploited~\cite{rodler2021}. 
We harden the five contracts against integer bugs with \toolname and compile the hardened source code to bytecode using the standard Solidity compiler. 
We replay and re-execute a total of \num{31216} transactions, of which 68 transactions are confirmed attacks.
\Cref{tab:patch-results} shows that \toolname effectively prevents all confirmed attack transactions---with a hardening effectiveness of \SI{100}{\percent} for integer bugs. 
Moreover, \toolname does not hamper the legitimate behavior of the studied contracts, i.e., all the benign transactions do not raise an alarm in \toolname, and their traces' effects are the same as the original one. 
Overall, \toolname introduces an average runtime gas overhead of \SI{1.64}{\percent}.

We now describe in detail the token contract belonging to the BeautyChain (BEC) token to demonstrate how \toolname hardens integer bugs. 
We select the BEC contract because it was actively exploited, resulting in a loss of $10^{58}$ BEC tokens~\cite{bec-report}. 
While the BEC contract is compatible with the ERC-20 standard for token transfers, it also implements custom functionality, such as the ability to transfer an amount of tokens to multiple receivers in its \emph{batchTransfer} function. 
\Cref{listing:bec-hard} shows the function that implements this behavior for the BEC token contract. 
Note that the highlighted lines (5 and 6) are added only after hardening through \toolname.
The \emph{batchTransfer} function takes an array of addresses (\emph{receivers}) and an integer as its input (\emph{value}). 
All addresses in the receivers array will receive an amount of tokens equal to \emph{value}. 
Interestingly, the developers were aware of integer bugs as they use the \emph{sub} and \emph{add} functions of the SafeMath library~\cite{safemath} in this function. 
However, the \emph{batchTransfer} function is vulnerable to an integer overflow due to a missing check before the multiplication in line 7, demonstrating that using SafeMath does not ensure full hardening against integer overflows.
Since both multiplication operands, i.e., the length of the array and \emph{value}, are inputs to this publicly callable function, an attacker can construct an exploit transaction by providing selected values for the parameters that trigger the multiplication overflow in line~7.
Consequently, the \emph{amount} variable overflows to a very low value or even to zero, which bypasses the check in line~9.
Due to the inserted overflow check in lines 5 and 6 of~\Cref{listing:bec-hard}, the multiplication overflow becomes unexploitable.

We further analyze the effectiveness of \toolname's integer bug hardening by extending our dataset to all contracts of the original OSIRIS dataset~\cite{Ferreira_Torres2018osiris}.
134 vulnerable contracts were deemed suitable for our experiment based on our criteria list described in~\Cref{sec:evaluation}.
To accurately perform this experiment, we need to ensure that OSIRIS does not report integer vulnerabilities for vulnerable code that is protected by \toolname's inserted hardening checks, because our implementation for integer vulnerabilities does not remove the vulnerability but rather inserts runtime checks that prevent exploitation.
Hence, we carefully sort out OSIRIS alarms for non-exploitable vulnerabilities.
Our evaluation results show that none of hardened contracts raised an alarm, which demonstrates the effectiveness of automated integer bug hardening with \toolname.

\paragraph{Reentrancy Hardening.}
For this experiment, we selected 5~contracts from the Sereum dataset~\cite{sereum} meeting the following conditions:
\begin{inparaenum}[(a)]
    \item source code is available,
    \item the contract uses a HCC-compatible solidity version, and
    \item we manually identified at least one true attack transaction, i.e., the contract is not a false positive.
\end{inparaenum}
Additionally, we included all contracts susceptible to reentrancy from the Ever-Evolving Game~\cite{yinzhi2020everevolvinggame} dataset.
We hardened the contracts against reentrancy bugs with \toolname and compile the hardened source code to bytecode using the standard Solidity compiler.

We replayed and re-executed \num{109} transactions, including \num{54} confirmed attack transactions.
\toolname's hardening introduces an average gas overhead of \SI{11.95}{\percent}.
\Cref{tab:patch-results} shows that \toolname prevents \emph{all} confirmed attack transactions, achieving \SI{100}{\percent} patch effectiveness for reentrancy hardening, while all the benign transactions do not raise any alarm in \toolname.
In contrast to the EEG dataset, we discovered two new reentrant transactions (contract 0x85d2\ldots{ea0b} and 0x516d\ldots{29a9}).
We analyzed the transactions and confirmed that those are indeed reentrant calls correctly prevented by \toolname.

\subsection{Practicality}
\label{sec:sguard-comparison}
\noindent
We compared \toolname coverage and effectiveness to the \sguard~\cite{nguyen2021-sguard} state-of-the-art source-to-source compiler to demonstrate the high practicality of \toolname.
To this end, we conducted a large-scale experiment on \num{10000} contracts sampled from \href{https://etherscan.io/}{etherscan.io} without any filtering using a timeout of one minute per contract and 64GB memory budget.
Note that a timeout of one minute is sufficient even for complex real-world contracts, as the compilation times in~\Cref{tab:top-results,tab:patch-results} show.
\toolname successfully processes \num{7174} (\num{71.74}\%), while \sguard processes only \num{3286} (\num{32.86}\%).

One might think that a timeout of one minute is too short for symbolic execution-based approaches.
Hence, we additionally compiled the 15~contracts containing known integer overflow and reentrancy vulnerabilities as well as the Top-25 contracts with \sguard using a timeout of \num{12} hours per contract.
Note that a timeout of 12 hours for all \num{10000} contracts is not possible as it would require months to complete the experiment.
Even with this excessive timeout, only 17 of the 40 contracts are compiled by \sguard, whereas \toolname compiles 36 out of 40~contracts.
In particular, \sguard processes only 4 of the Top-25 contracts successfully, while \toolname is able to successfully compile 20 contracts.
In addition, \sguard does not add any patch to any contract of our integer bug dataset, including the real-world ERC20 Token contracts that were actively exploited in 2018.
\begin{table}[tbp]
    \centering
    {
      \begin{adjustbox}{max width =\linewidth}\begin{tabular}{@{}l|@{ }r@{ }|rr|r@{ }r@{ }|@{}r@{}}
            \toprule
            \multirow{3}{*}{Contract} & \multirow{3}{*}{Checks} & \multicolumn{2}{c|}{\# Transactions}   & \multicolumn{2}{c|}{Gas Overhead} & \multicolumn{1}{@{ }c}{Compile}\\\cline{3-4}\cline{5-6}
                                      &                             & \multirow{2}{*}{Total} & \multirow{2}{*}{Attacks} & \multicolumn{1}{@{ }c}{mean} & \multicolumn{1}{@{ }c|}{median}  & \multicolumn{1}{@{ }c}{Time}\\
                                      & & &  & \multicolumn{1}{@{ }c}{(\%)} & \multicolumn{1}{@{ }c|}{(gas)} & \multicolumn{1}{@{ }c}{(s)} \\
            \midrule
            BEC                       & 5        & \num{9296}  & \num{1}                 & \SI{0.43}{\percent}                  & \num{103}         & \num{4.04}\\
            HXG                       & 8        & \num{1284}  & \num{10}                & \SI{4.37}{\percent}                  & \num{2884}        & \num{4.36}\\
            SCA                       & 31       & \num{286}   & \num{1}                 & \SI{3.32}{\percent}                  & \num{1590}        & \num{5.20}\\
            SMT                       & 14       & \num{9816}  & \num{1}                 & \SI{4.36}{\percent}                  & \num{1743}        & \num{3.93}\\
            UET                       & 14       & \num{10534} & \num{55}                & \SI{3.41}{\percent}                  & \num{1571}        & \num{4.23}\\
            \midrule
            Total/\textbf{Average}   & 72       & \num{31216} & \num{68}                & \textbf{3.16 \%}           & \textbf{1579} &  \textbf{4.35}\\
            \midrule
            \midrule
            \texttt{0x0eb\ldots{43e}}         & 1        & \num{2}   & \num{1}                   & \SI{1.32}{\percent}                   & \num{551}         & \num{4.54}\\
            \texttt{0x2c4\ldots{e9d}}         & 5        & \num{1}   & \num{1}                   & -                     & -                 & \num{5.48}\\
            \texttt{0x72f\ldots{82e}}         & 3        & \num{2}   & \num{1}                   & \SI{0.17}{\percent}                   & \num{408}         & \num{6.29}\\
            \texttt{0x85d\ldots{a0b}}         & 5        & \num{26}  & \num{26}                  & -                       & -                 & \num{5.80}\\ 
            \texttt{0x516\ldots{9a9}}         & 5        & \num{8}   & \num{8}                   & -                       & -                 & \num{5.90}\\ 
            \texttt{0xd4c\ldots{69e}}         & 2        & \num{8}   & \num{1}                   & \SI{3.02}{\percent}                  & \num{642}         & \num{6.48}\\
            \texttt{0x4a8\ldots{a38}}         & 1        & \num{43}  & \num{10}                  & \SI{11.00}{\percent}                 & \num{27}          & \num{6.85}\\
            \texttt{0xaae\ldots{0b8}}         & 1        & \num{8}   & \num{2}                   & \SI{0.38}{\percent}                  & \num{424}         & \num{5.01}\\ 
            \texttt{0xb4c\ldots{999}}         & 1        & \num{7}   & \num{3}                   & \SI{47.46}{\percent}                 & \num{10918}       & \num{4.71}\\ 
            \texttt{0xb7c\ldots{4f7}}         & 1        & \num{4}   & \num{1}                   & \SI{20.31}{\percent}                 & \num{942}         & \num{4.21}\\ 
            \midrule
            Total/\textbf{Average}   & 25       & \num{109} & \num{54}                  & \textbf{11.95 \%}      & \textbf{1987} & \textbf{5.52}\\
            \bottomrule
        \end{tabular}
        \end{adjustbox}
    }
    \vspace{1ex}
    \caption{HCC effectiveness for ERC-20 tokens (upper part) and reentrancy (lower part).}
    \label{tab:patch-results}
\end{table}
We also observed that existing approaches hamper legitimate transactions (false positive) when checking for reentrancy vulnerabilities.
For instance, \sguard applies a locking mechanism similar to OpenZeppelin's \emph{ReentrancyGuard}~\cite{openzeppelin-reguard}, i.e., it creates a single mutex-lock for the whole contract.
As mentioned in \Cref{sec:re-hardening}, this locking strategy prevents legitimate access patterns, like withdrawals from different accounts within the same transaction, that cannot lead to state-inconsistency.
In contrast, \toolname's locking strategy is field- and offset-sensitive as it considers both, fields of data structures and indices to mappings without limits to nesting, thereby allowing these access patterns.

\subsection{Other Runtime Metrics}
\label{sec:morevulns-eval}

Similar to reentrancy and integer bug hardening, we also evaluate the gas overhead of \toolname's other mitigations (\Cref{sec:suicidal} to~\ref{sec:uncheckedcalls}) in a separate experiment.
As there are very few real-world transactions that trigger these vulnerabilities and their functions, we perform this experiment based on the code samples in \Cref{lst:suicidal} to~\ref{lst:unchecked}.

\begin{wraptable}{O}{.7\linewidth}
    \centering
\begin{tabular}{l|r|r}
    \toprule
    \multicolumn{1}{@{ }c|}{Vulnerability} & \multicolumn{1}{@{ }c|}{Deployment} & \multicolumn{1}{@{ }c}{Execution } \\
     & \multicolumn{1}{@{ }c|}{Overhead} & \multicolumn{1}{@{ }c}{Overhead} \\
     \midrule
     Suicidal                           &  107.9\%    &   42.0\%   \\
     \texttt{tx.origin}                 &    0.0\%    &    0.0\%   \\
     Untrusted \textit{delegatecall}    &   14.6\%    &    1.6\%   \\
     Unchecked low-level call           &    2.7\%    &    0.2\%   \\
     \midrule
     \textbf{Average}                   &   31.3\%    &   10.9\%   \\
     \bottomrule
    \end{tabular}
    \vspace{1ex}
    \caption{\centering Gas overhead introduced by \toolname's hardening patches.}%
    \label{table:morevulns}
\end{wraptable}

As shown in \Cref{table:morevulns}, the execution and deployment overheads with \toolname's hardening code of all contracts are negligible, except for the contract from \Cref{lst:suicidal}. This outlier is due to the fact that the additional runtime check doubles the code size of the function, as otherwise the \textit{destroy} function would only consist of the function call to the \textit{selfdestruct} function, which is very cheap in terms of gas costs.
Similarly, the hardened version of \Cref{lst:suicidal} introduces a larger code size and deployment overhead due the insertion of the constructor and the \textit{owner} variable.
Note that deployment overhead is negligible, as it is a one-time cost.

%% file: sections/fabric.tex
\noindent
Hyperledger Fabric~\cite{fabric} is one of the most popular permissioned blockchain platforms, which supports Golang-based smart contracts. 
To apply \toolname to Hyperledger Fabric, we apply the following adaptations.
Since not all compilers generate an AST in a machine-readable format, and in order to reduce the differences in ASTs generated by language-specific compilers (such as the Golang compiler for Golang code), we make use of a generic parser generator such as ANTLR~\cite{antlr}.
ANTLR can generate AST models for many existing programming languages, such as Golang. 
As ANLTR provides a uniform parsing behavior across all the supported languages, it ensures minimal modifications to the CPG enrichment phase, as we do not need to account for compiler-specific behavior for each new programming language.

The code generation phase requires minimal adaptation to emit valid code in the programming language considered, Golang in this case. Since the logical operators are usually similar across languages, we only need to change the error reporting mechanism. This requires changing the $assert\left(condition\right)$ used in Solidity to the standard "if not condition, return error" of Golang as follows:

\begin{lstlisting}[language=go,frame=none,numbers=none]
if(!condition) {
  return nil, errors.New("HCC: Illegal Operation") }
\end{lstlisting}

Lastly, we modify the CPG enrichment phase to match platform-specific APIs to their existing equivalent (of Ethereum or any other already-integrated platform). Indeed, different smart contract platforms each provide their own specific APIs on how to interact with other smart contracts and how to access the storage system. 
In the case of Fabric, this includes Fabric's storage access patterns (i.e., the \emph{PutState} and \emph{GetState} methods) and external call patterns (Fabric's \emph{InvokeChaincode} method).

We note that, when incorporating Fabric support into \toolname, only 101 lines of code were added, and 8 lines of code were modified (excluding the code associated with the AST generation step). These represent approximately 2.75\%, of the overall 3953 lines of code.

Note that, while reentrancy attacks are specific to Ethereum, other types of attacks, such as integer overflows and underflows, are directly applicable to Fabric's smart contracts.
This also means that the bug discovery and hardening phases for integer bugs (c.f., \Cref{sec:io-hardening}) can be reused without any modifications.
Owing to Golang's type inference, it is not always easy to differentiate between signed and unsigned integers in the AST, as type information is not always available.
For this reason, when \toolname cannot ensure with certainty the type, it always assumes that the type is signed as it creates additional checks compared to the unsigned version, ensuring the effectiveness of the hardening regardless of whether the actual type (although part of the check may be superfluous if the type was unsigned).

\begin{wraptable}{O}{.7\linewidth}
    \centering
\scalebox{0.85}{\begin{tabular}{l|c|c|c}
    \toprule
    \textbf{Contract} & \textbf{LOC Overhead} & \textbf{Execution } & \textbf{Execution}  \\
     &  & \textbf{Overhead} & \textbf{Overhead} (ns) \\
     \midrule
     BEC& 14\% &  2.8\% & 4\\
     HXG&  38\%&   30.2\%& 78 \\
     SCA &  34\%&   29.8\%& 151\\
     SMT &  36\%&   28.2\%& 96\\
     UET&  33\%&   26.8\%&52\\
     \midrule
     \textbf{Average}& 31\% &23.5\%&76.2\\
     \bottomrule
    \end{tabular}}
    \vspace{1ex}
    \caption{\centering Execution overhead introduced by \toolname in Hyperledger Fabric.}%
    \label{table:fabric}
\end{wraptable}

\paragraph{Evaluation of \toolname on Fabric.}
We evaluate our integration of \toolname with Fabric using the five ERC-20 token contracts from the Osiris~\cite{Ferreira_Torres2018osiris} dataset.
To do so, we translate vulnerable functions from Solidity to Golang and then apply \toolname on the resulting smart contract.
We then measure the overhead introduced (in LOC and execution time) in each function by \toolname.
\Cref{table:fabric}~shows our results; here each point is averaged over 1000 independent runs.
Our results show that \toolname introduces an average code overhead of 31\%; this corresponds to adding the 3 lines required for each assert introduced by \toolname.
On the other hand, \toolname introduces an average of 23.5\% in terms of execution times (around 76 nanoseconds).
Note that this overhead varies between BEC and the other considered contracts.
This is mainly due to the fact that the BEC contract has only one patch which introduces little overhead.
Recall that the Fabric framework requires a few seconds to process a transaction---as such, an overhead of hundreds of nanoseconds can be easily tolerated.

Beyond Hyperledger Fabric, we argue that the reliance on ANTLR ensures support for many smart contract based languages since ANTLR has grammars targeted at many popular programming languages, such as Java, C++, C\#, Python.
This means that \toolname can support a wide variety of blockchain platform such as Hyperledger Sawtooth (Python)~\cite{sawtooth} and Neo (Java, Python, C\#)~\cite{neo-project} without incurring significant changes in the AST generation step.

%% file: sections/conclusion.tex
\noindent While the popularity and adoption of blockchain technologies steadily increase, little attention is paid to developing secure contracts. \toolname is the first practical language-independent approach that keeps the developer out of the loop, hardening smart contracts against a variety of attacks. It introduces a new and generic smart contract representation, namely a code-property graph, which allows modeling of diverse vulnerabilities in Ethereum and Hyperledger Fabric thereby addressing the high dynamics in this application domain.
We demonstrate that \toolname is applicable to complex and high-value state-of-the-art contracts in Ethereum preventing all known reentrancy and integer bugs.

%% file: sections/acknowledgements.tex
\subsubsection{Acknowledgements}
This work has been partially funded by the Deutsche Forschungsgemeinschaft (DFG, German Research Foundation)---SFB~1119 (CROSSING) 236615297 within project~T1, EXC 2092 (CASA) 39078197---and the European Union ERC CONSEC under grant No.~101042266 and ECH2020 TeraFlow Project under grant agreement No. 101015857.
The views and opinions expressed are those of the authors only and do not necessarily reflect those of the European Union or the European Research Council Executive Agency.
Neither the European Union nor the granting authority can be held responsible for them.